\documentclass[lettersize,journal]{IEEEtran}
\usepackage{amsmath,amsfonts}
\usepackage{algorithm}
\usepackage{algpseudocode}
\usepackage{array}
\usepackage[caption=false,font=footnotesize]{subfig}
\usepackage{textcomp}
\usepackage{stfloats}
\usepackage{url}
\usepackage{verbatim}
\usepackage{graphicx}
\usepackage{multirow}
\usepackage[utf8]{inputenc}
\usepackage{cite}
\usepackage{makecell}
\usepackage{diagbox}
\usepackage{placeins}
\begin{document}
\title{Towards Sustainable Federated Learning via Carbon-Aware Client Selection}

\author{
    Wentao~Gao,
    Omid~Tavallaie,~\IEEEmembership{Member,~IEEE},
    Albert~Y.~Zomaya,~\IEEEmembership{Fellow,~IEEE}
    \thanks{Wentao Gao and Albert Y. Zomaya are with school of computer science, The University of Sydney, Australia. Email: wentao.gao@sydney.edu.au, albert.zomaya@sydney.edu.au}
    \thanks{Omid Tavallaie is with Department of Engineering Science, University of Oxford and Department of Computer Science, University of Western Australia. Email: omid.tavallaie@uwa.edu.au, omid.tavallaie@eng.ox.ac.uk}
}

\markboth{}%
{Shell \MakeLowercase{\textit{et al.}}: A Sample Article Using IEEEtran.cls for IEEE Journals}


\maketitle

\begin{abstract}
The Information/Communication Technology (ICT) industry is severely challenging the global environment due to the sharply rising carbon emissions generated by machine learning (ML) workloads. Federated Learning (FL) is one of the prominent distributed ML systems used in applications with billions of client devices, such as Google Gboard and Apple Siri. Nearly half of the total carbon footprint in FL originates from client-side computation during local training. However, existing client selection strategies in FL primarily focus on minimizing energy consumption rather than carbon emissions, despite the fact that carbon impact depends not only on computational load but also on the underlying carbon intensity of the energy source. In this paper, we address this gap by proposing a carbon-aware fair client selection strategy (CA-FCS) designed to minimize the carbon emissions associated with model training on client devices. In contrast to existing approaches for reducing carbon emissions in FL, we incorporate hourly carbon intensity and a realistic energy consumption model developed using empirical measurements from Google Pixel smartphones to estimate client-side carbon emissions in FL. Since repeatedly selecting low-carbon clients may reduce data diversity and degrade global model performance, particularly under non-independent and identically distributed (non-IID) data settings, CA-FCS incorporates fairness into client selection to mitigate this issue. We conduct extensive experiments under realistic FL settings for mobile devices, where client participation is constrained by device availability conditions, such as charging status and Wi-Fi connectivity. The results demonstrate that our method reduces client-side carbon emissions by up to 46\% while maintaining competitive model accuracy.

\end{abstract}


\section{Introduction}\label{sec:intro}
\IEEEPARstart{T}{he} challenge from climate change, driven by carbon emissions from human activities, remains profoundly severe. According to \cite{wri_state_of_climate_action_2025}, the pace and scale of global efforts to reduce greenhouse gas emissions and enhance carbon removals are insufficient to achieve the temperature control target set by the Paris Agreement. The Information/Communication Technology (ICT) industry has emerged as one of the largest consumers of electricity \cite{towardcarbonaware}. The annual greenhouse gas emissions from the ICT ecosystem now account for over 2\% of global emissions \cite{carbonfootprintoffl}. With the large-scale deployment of artificial intelligence, the carbon footprint of machine learning (ML) applications has risen sharply and become a significant part of the ICT carbon footprint \cite{sustainableAI1}, \cite{sustainableAI2}. Among various applications of ML, Federated Learning (FL) stands out as a particularly important distributed ML system because it serves billions of client devices in real-world deployments, such as Google Gboard and Apple Siri. In a typical FL process, clients perform local training on their private datasets and periodically upload model updates to the server for model aggregation. Recent research by Meta \cite{greenfl} indicates that concerns about the carbon footprint of large-scale FL are growing. Their experiment, which was performed on over a million clients, shows that around 50\% of carbon emissions in the FL process come from client-side computing, and another approximately 50\% come from the communication between clients and the server.

\begin{figure*}
    \centering
    \subfloat[Random Client Selection]{
    \centering
        \includegraphics[width=90 mm, height=45 mm]{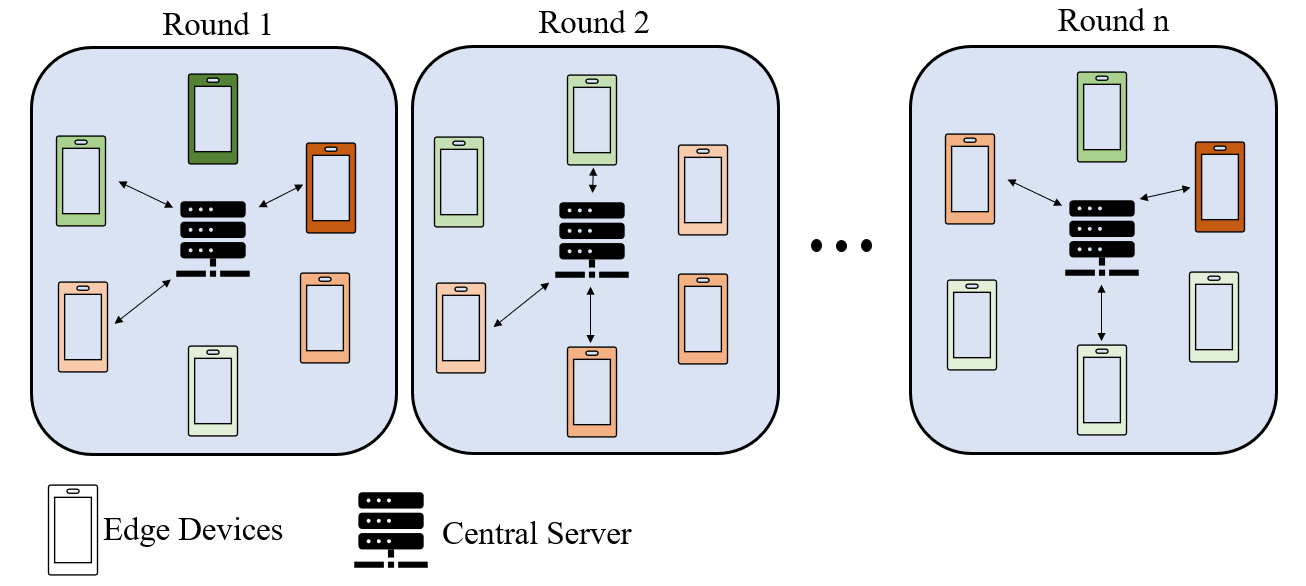}
        \label{fig: Vanilla FL}
    }
    \subfloat[Carbon-aware Client Selection]{
    \centering
        \includegraphics[width=90 mm, height=45 mm]{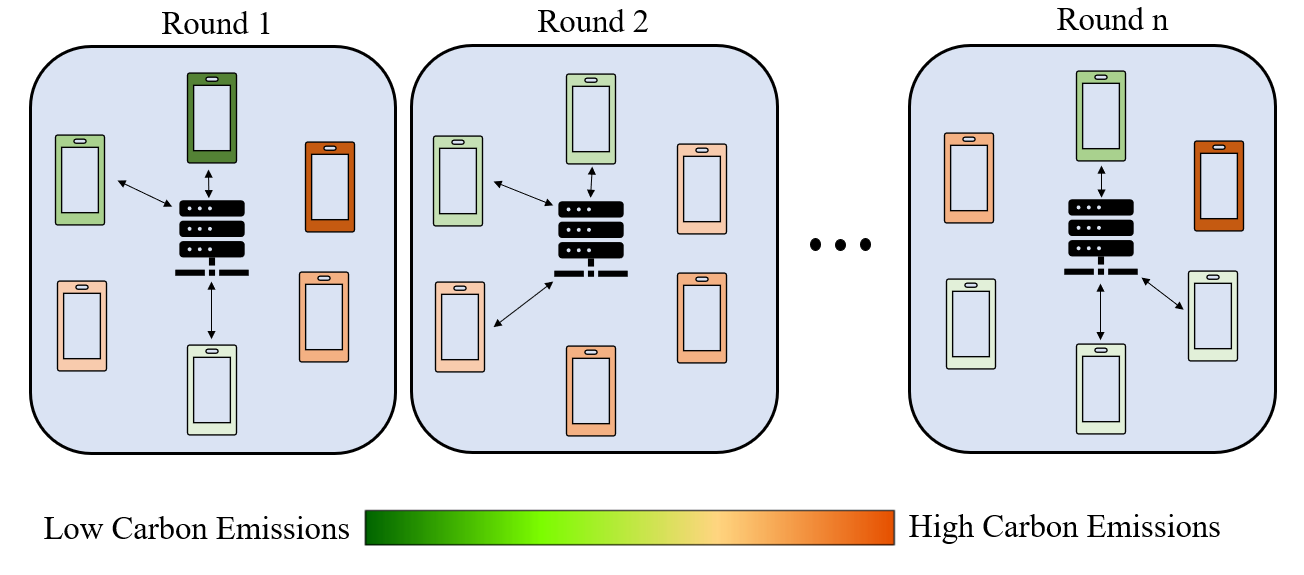}
        \label{fig: Carbon-aware FL}
    }
    \caption{Random vs. Carbon-Aware Client Selection. (a) Random client selection: In each training round, a fixed number of clients are randomly selected to participate in FL. (b) Carbon-aware client selection: In each training round, a fixed number of clients with the lowest estimated carbon emissions are selected to participate in FL.}
    \label{fig:clientSelection}
\end{figure*}

As FL applications typically operate at a large scale with numerous clients, employing client selection algorithms to determine which clients participate in training constitutes a crucial component of the overall process. One of the most popular client selection methods is random client selection, which selects clients randomly in each training round \cite{fedavg}. Many state-of-the-art proposed client selection strategies focus on improving the performance of FL, such as accuracy and convergence speed \cite{poc}, \cite{oort}. However, attention to the environmental impact of client-side computation is limited. The carbon emissions from local training of a client are given by the product of the energy consumption and the carbon intensity during its local training. Carbon intensity refers to the amount of carbon emissions produced per unit of energy production. It varies across locations and over time due to differences in energy resources and temporal fluctuations in the power system \cite{carbonintensity}. For example, a region that relies on solar tends to exhibit lower carbon intensity compared to a region using coal. Even in a single day, at the same location, carbon intensity may vary across different time periods. Although there are some proposed energy-efficiency client selection strategies (such as \cite{eafl}), it is worth noting that energy-efficiency and carbon-efficiency are not equivalent. FL typically requires clients (such as mobile devices) to participate only when they are connected to Wi-Fi and charging to avoid accidentally dropping out \cite{greenfl}. It results in heterogeneous participation times across clients. Consequently, even within the same geographic region, clients may have varying carbon intensities during local training due to differences in the time when they implement local training, leading to disparities in their associated carbon emissions. This motivates the need for a carbon-aware client-selection strategy for FL with mobile devices as clients, which can reduce the carbon emissions from client-side computing by prioritizing clients with lower carbon emissions from local training. Fig.\ref{fig:clientSelection} illustrates an example of client selection outcomes using random selection and selection aimed at minimising carbon emitted during training.

Exclusively selecting low-emissions clients may lead to biased client selection and degrade the performance of the aggregated global model, particularly with non-independent and identically distributed (non-IID) data settings. As FL selects only a fraction of clients per round for training, employing a fair client selection strategy is essential. Repeatedly selecting certain clients while excluding others can introduce bias and degrade model performance \cite{rafi2024fairness}. In this paper, we propose a carbon-aware fair client selection strategy (CA-FCS) for FL with mobile devices as clients. In the design of CA-FCS, to preserve performance for any data distribution (including the extreme case of non-IID), in addition to estimating and taking the carbon emissions of each client into account, we consider two conditions to ensure fairness in the client selection process: 1) by limiting the minimum and maximum number of times that a client can be selected for model training, and 2) by preventing selection of the same client in 2 consecutive training rounds. Besides, in CA-FCS, the precise carbon emissions for a client are estimated without sharing private information (such as geographical position) with the server. The estimation is based on the energy consumption information (such as voltage) and the fine-grained carbon intensity at the time when the client is usually plugged into a power outlet and connected to Wi-Fi. The main contributions of our work are summarized as follows:
\begin{itemize}
\item 1) By considering the varied carbon intensity and power model of mobile devices, we propose CA-FCS, a fair carbon-aware client selection method for FL that dynamically adjusts the set of client devices participating in model training based on their estimated carbon emissions. To the best of our knowledge, CA-FCS is the first client selection method proposed for FL that employs fine-grained carbon intensity to estimate the carbon emissions from client devices and further guides the selection process.

\item 2) To avoid sacrificing model performance (accuracy) while minimizing carbon emissions by using client selection, CA-FCS considers not only the estimated carbon footprint of local model training at each client, but also the fairness of the client selection process. This prevents repeatedly selecting low-carbon clients. As a result, CA-FCS provides a fair client selection strategy that maintains a competitive number of training rounds and accuracy under both IID and non-IID settings. 

\item 3) To estimate the precise carbon emissions of model training on a client device, CA-FCS proposes a realistic solution that was implemented and tested on Google Pixel phones with different hardware specifications. By using the electric charge consumed for local training, the voltage of the edge device's battery, and the fine-grained carbon intensity, the client can estimate the carbon emissions of its local training accurately without sharing private information (such as location) with the server. This approach is different from solutions previously proposed in the literature that incorrectly assume minimizing power consumption is always equal to reducing carbon emissions.
\end{itemize}

The remainder of this paper is organized as follows: Background is introduced in Section II. Section III discusses related works. The key challenges of employing a carbon-aware client selection strategy are explained in Section IV. In Section V, we discuss the measurement of carbon emissions on client devices. CA-FCS is proposed in  Section VI. Section VII explains how we implement the experiments of CA-FCS on Google Pixel phones. The experimental results are evaluated in Section VIII. In Section IX, we discuss addressing the limitations and future work of our method. Finally, Section X concludes the paper.

\section{Background}
FL \cite{fedavg} is a privacy-aware distributed ML framework that enables multiple clients to collaboratively optimize a shared global model under the coordination of a central server while keeping raw training data localized on client devices. In large-scale FL systems, the number of participating clients can reach billions, making it infeasible for all clients to participate in every training round due to constraints such as limited bandwidth, computational resources, and energy availability. Therefore, in each training round, only a subset of clients is selected. A key challenge in FL is the non-IID distribution of data across clients, which can degrade model accuracy and slow convergence \cite{non-iid,personalizefl}, while also raising concerns about fairness in client selection. Consistently excluding clients with unique data may prevent the global model from capturing important data distributions.

Recently, increasing attention has been given to the environmental impact of machine learning, leading to the emergence of sustainable ML research \cite{alswaitti2025training}, \cite{iftikhar2025enhancing}, \cite{sustainableAI3}. The ICT industry consumes 2-3\% of global electricity today, predicted to reach 8-21\% by 2030 \cite{towardcarbonaware}. In large-scale FL, a substantial amount of ML tasks are performed on client devices. The carbon emissions from the server-side computation in FL are only 1-2\%, while client-side computation contributes 46-50\%. Upload and download networking produce 27-29\% and 22-24\%, respectively \cite{greenfl}. 

The carbon emissions from a mobile device's local training can be estimated as the product of energy consumption and carbon intensity, where carbon intensity is a measure of how much CO$_2$ is emitted per unit of energy produced. Carbon intensity exhibits spatial and temporal heterogeneity. Regions relying more on renewable or low-carbon sources (e.g., wind, solar, and nuclear power) generally have lower carbon intensity \cite{few2024geographically}, and even within the same region it varies over the day with electricity demand and the generation mix \cite{zivin2014spatial}. Since clients typically participate in FL only when connected to stable networks and power sources \cite{kairouz2021advances}, even clients in the same region may experience different carbon intensities during training (discussed in Section \ref{sec: carbon emissions measurement}). Different from centralized training, which can be scheduled in low-carbon regions and time periods, FL inherently involves geographically distributed and temporally constrained client devices, making it particularly challenging to optimize carbon efficiency.

\section{Related Works}
Random Client Selection (Fig.\ref{fig:clientSelection}), as adopted in FedAvg \cite{fedavg}, randomly samples a subset of clients in each training round. However, it does not
consider sustainability-related factors. Strategies such as Oort \cite{oort} and PoC \cite{poc} improve accuracy and convergence efficiency, but primarily optimize statistical objectives while ignoring the
environmental impact of FL. Recently, a number of energy-efficient client selection approaches have been proposed \cite{eafl}, \cite{fedecs}, and \cite{albelaihi2022green}. These methods reduce the energy consumption of local training and mitigate client dropouts due to low battery levels, but energy efficiency is not equivalent to carbon
efficiency. The carbon emissions of client devices depend not only on the energy consumption but also on the corresponding carbon intensity of the electricity, which varies across both location and time. Recent studies (\cite{cefl} and \cite{cafe}) incorporate carbon intensity into the selection process to improve environmental sustainability. However, they rely on coarse-grained carbon intensity estimates, such as annual or monthly averages. Such simplifications may not accurately reflect real-world conditions, as carbon intensity exhibits significant temporal variation even within the same geographic location on the same day \cite{khan2018analysis}, which leads to inaccurate emissions estimates and suboptimal selection decisions. \textit{To the best of our knowledge, \cite{cefl} and \cite{eafl} are the only existing methods that can be applied to minimize carbon emissions as an optimization objective with fine-grained carbon intensity.} They also overlook the differences in energy consumption caused by hardware heterogeneity \cite{gao2024federated} (e.g., processors, memory architectures, and battery systems), even when clients train in the same location at the same time \cite{esquivel2024performance}. Furthermore, they may lack explicit fairness constraints. So, repeatedly prioritizing low-emission clients causes participation imbalance and reduces data diversity, particularly under highly non-IID settings \cite{fairness}. Moreover, existing carbon-efficiency client selection strategies may lead to privacy concerns. The server typically determines the carbon efficiency of clients based on their carbon intensity by asking them to provide carbon intensity or geographical location. As a result, the server can easily infer clients' geographical locations, since carbon intensity is spatially heterogeneous \cite{aryai2024real}. In parallel, some studies have leveraged the temporal variability of carbon intensity to reduce carbon emissions from client-side computing through scheduling in FL. These methods schedule clients' local training processes during periods of the day when the carbon intensity is low, such as \cite{arputharaj2025green} and \cite{yang2025survey}. However, the effectiveness of such approaches may diminish when the clients are mobile devices. Because mobile devices, such as smartphones, are typically required to be both charging and connected to Wi-Fi before they are allowed to participate in FL \cite{greenfl}. Consequently, clients are unable to freely determine the timing of their participation in FL.

\section{Key Challenges of Employing a Carbon-aware Client Selection Strategy}
Employing a carbon-aware client selection strategy in practical FL with mobile devices faces three key challenges: precise estimation of the carbon emissions from a mobile device in a practical environment, balancing the sustainability and the performance of the global model in FL, and privacy concerns of sharing clients' information for estimating the carbon emissions. In this section, we discuss each of these challenges in detail.

\textit{Precise estimation of the carbon emissions from a mobile device in a practical environment}: Carbon emissions from a mobile device's local training are jointly determined by the energy consumption of local training and the carbon intensity of the energy supply at the position of the device during the training period. Both factors are heterogeneous in practice. Energy consumption varies across clients due to hardware heterogeneity, so estimating it solely from reported battery percentage is inherently inaccurate. For example, two mobile devices with battery capacities of 5000 $mAh$ and 3000 $mAh$ experiencing the same battery level reduction from 100\% to 75\% consume substantially different amounts of electrical charge. Battery level changes reflect not only the energy consumed by FL training but also the energy used by background processes and other concurrent system activities. Carbon intensity varies significantly across both geographic locations and time periods. Because the sustainability of energy sources is different across regions, the carbon intensity may differ significantly between locations. Even in the same place, electricity generation mixes may change dynamically throughout the day due to fluctuations in renewable energy availability and electricity demand. In practical FL deployments, client availability varies throughout the day, as users charge their devices and connect to Wi-Fi at different times, resulting in different participation windows across clients. Consequently, coarse-grained carbon intensity, such as daily average carbon intensity, cannot accurately represent the actual carbon intensity associated with a device during local training. For instance, if a mobile device performs local training at 3:00 a.m. on January 1, 2024, in New South Wales, Australia, the carbon intensity at that time is 299.58 $gCO_{2}/kWh$, substantially lower than the daily average of 493.04 $gCO_{2}/kWh$. This discrepancy highlights how relying on daily average carbon intensity can lead to inaccurate estimates of the carbon emissions associated with local training. \cite{electricitymaps}.

\textit{Balancing the sustainability and the performance of the global model in FL}: Excessively selecting clients with low carbon emissions can apparently reduce the carbon footprint from client-side computing. However, rarely or never selecting clients with high carbon emissions may lead to a degradation in the performance of the global model, such as lower accuracy or longer time for convergence. Since clients' local datasets are usually heterogeneous, insufficient participation from certain clients can prevent the global model from capturing features that exist only in those clients' datasets. This issue becomes more severe in practical FL scenarios where data distributions across clients are highly skewed (non-IID). 

\textit{Privacy concerns of sharing clients' information for estimating carbon emissions}: Carbon-aware client selection methods should be designed to minimize the amount of information required from clients for estimating their carbon emissions. Ideally, such methods should rely on minimal information, preferably information that does not reveal clients' private details or behavioral patterns. For instance, requiring clients to share their geographical locations, charging schedules, or Wi-Fi connectivity patterns may introduce additional privacy risks. Therefore, an important research direction is to develop carbon-aware client selection methods that can accurately estimate carbon emissions while requiring minimal information disclosure from clients. More importantly, it should not be overlooked that one of the primary motivations behind FL is to preserve the privacy of clients (users). This fundamental objective must not be compromised for the sake of any optimization, including reducing carbon emissions.

\section{Estimating Carbon Emissions for Clients}
This section introduces the concept of carbon intensity and presents the formula for estimating carbon emissions from mobile devices. It then explains how CA-FCS estimates these emissions while preserving client privacy by avoiding the disclosure of sensitive information, such as client location. Table. \ref{tab:notation} shows the main notations used in this paper.

\begin{table}[b]
    \caption{Declaration of Notations}
    \begin{tabular}{|p{8 mm}|p{8 cm}|}
    \hline
    Notation & Definition \\
    \hline
    \(S\) & The set of all clients \\
    \hline
    \(K_{min}\) & The minimum number of times that a client can be selected \\
    \hline
    \(K_{max}\) & The maximum number of times that a client can be selected\\
    \hline
    \(N\) & The total number of clients \\
    \hline
    \(x\) & The required number of clients in each FL training round \\
    \hline
    \(T_{prep}\) & The required number of initial training rounds \\
    \hline
    \(E_i\) & The energy consumption of local training on client \(i\) \\
    \hline
    \(Q_i\) & The electric charge consumed for local training on client \(i\) \\
    \hline
    \(V_i\) & The device voltage of client \(i\) \\
    \hline
    \(C_i^t\) & The carbon emissions of local training on client \(i\) at round \(t\) \\
    \hline
    \(CI_i^t\) & The carbon intensity of client \(i\) at round \(t\) \\
    \hline
    \(K_i\) & The number of times that client \(i\) has been selected \\
    \hline
    \(S_t\) & The set of selected clients at round \(t\), where \(t \in [0,T_{prep}-1]\) \\
    \hline
    \end{tabular}
    \label{tab:notation}
\end{table}

\label{sec: carbon emissions measurement}
\subsection{Carbon Intensity}
\begin{figure}
\centering
\includegraphics[width=75 mm, height=48 mm]{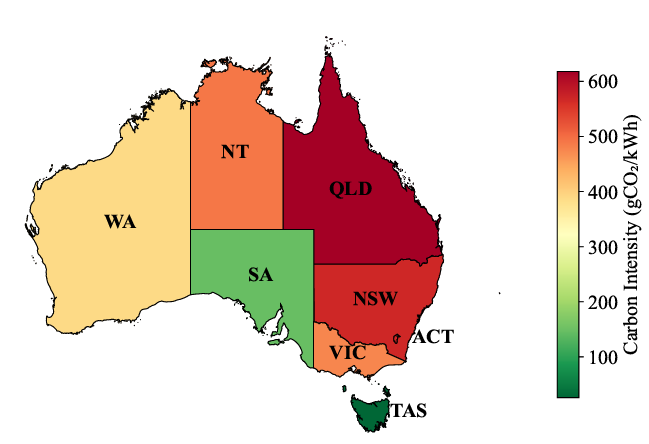}
\caption{Carbon intensity ($gCO_2/kWh$) of Australian states on 19th Oct 2026 7:00 p.m. Australia Eastern Time. The carbon intensity of Queensland (QLD) is highest, and the carbon intensity in South Australia (SA) and Tasmania (TAS) is apparently lower than that of other states.}
\label{fig: Australia Carbon Intensity Map}
\end{figure}

Carbon intensity refers to the amount of CO$_2$ equivalent emitted per unit of energy \cite{carbonintensity}. In this study, it is defined as the grams of  CO$_2$ equivalent emitted per kilowatt-hour ($kWh$) of electricity generated ($gCO_{2}/kWh$). A lower carbon intensity reflects a cleaner energy source. It varies across regions due to differences in energy generation structures and resource availability. Regions that rely heavily on fossil fuels typically exhibit higher carbon intensity, as these sources emit large amounts of CO$_2$ per unit of energy produced. In contrast, areas with a higher penetration of renewable or green energy sources such as hydroelectric, solar, or wind power tend to achieve significantly lower carbon intensity levels. Consequently, the carbon intensity associated with electricity consumption on client devices is inherently location-dependent and should be carefully considered in the design of carbon-aware FL systems. To illustrate the spatial heterogeneity of carbon intensity, Fig. \ref{fig: Australia Carbon Intensity Map} presents the carbon intensity across different states in Australia at 7:00 p.m. on October 19, 2026. Among all states, Queensland (QLD) exhibits the highest carbon intensity, exceeding 600 $gCO_2/kWh$, whereas Tasmania (TAS) demonstrates a substantially lower level, which is below 25 $gCO_2/kWh$.

\begin{figure}
\centering
\includegraphics[width=75 mm, height=45 mm]{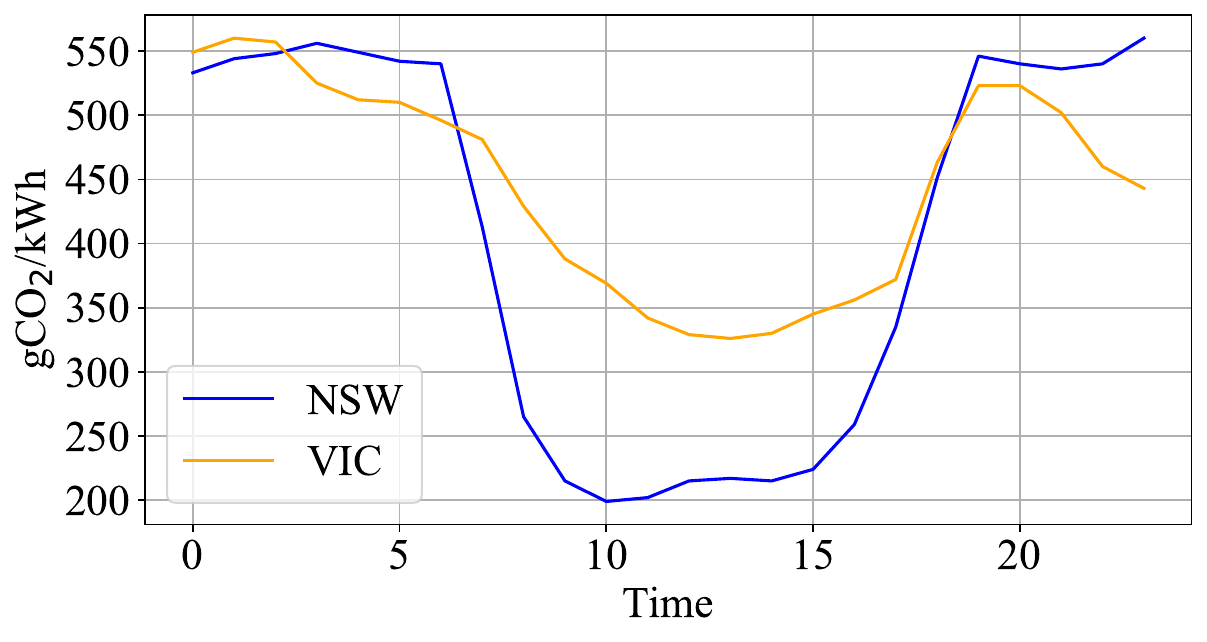}
\caption{The daily carbon intensity changes of NSW and VIC in Australia on 18/10/2026. In both NSW and VIC, carbon intensity varies over time. Significant differences in carbon intensity levels can be observed between daytime and nighttime.}
\label{fig: NSW vs VIC carbon intensity}
\end{figure}

In addition to spatial variation, carbon intensity exhibits significant temporal dynamics primarily driven by fluctuations in electricity demand and changes in the generation mix over time. For example, during the daytime, the availability of cleaner energy sources, such as solar power, may result in lower carbon intensity compared with nighttime. As a result, \textit{carbon intensity is inherently time-dependent and cannot be assumed to be constant}. Fig.\ref{fig: NSW vs VIC carbon intensity} illustrates the temporal variations of carbon intensity over a 24-hour period on October 18, 2026, in New South Wales (NSW) and Victoria (VIC), Australia. Both regions exhibit similar diurnal patterns: carbon intensity decreases in the early morning and rises during the nighttime.

\subsection{Estimating Carbon Emissions from Local Training on Mobile Devices}
The carbon emissions \(C_i^t\) (\(gCO_2\)) of client \(i\) in round \(t\) are estimated from the energy consumption \(E_i\) (\(kWh\)) for local training and the corresponding carbon intensity \(CI_i^t\) (\(gCO_2/kWh\)) \cite{hasan2025carbon}:
\begin{equation}
 C_i^t = E_i \cdot CI_i^t.
 \label{eq: carbon emissions calculation}
\end{equation}

The energy consumption \(E_i\) is the product of the electric charge consumed for local training ($mAh$) and the device voltage ($V$):
\begin{equation}
E_i = Q_i \cdot V_i,
 \label{eq: electricity consumption calculation}
\end{equation}
where \(Q_i\) and \(V_i\) are electric charge consumed and device voltage of client \(i\), respectively.

\subsection{The Estimation of Precise Carbon Emissions without Sharing Private Information in CA-FCS}
\label{sec: process of estimating carbon emissions in CA-FCS}
In CA-FCS, the server requires clients to independently perform the estimation of carbon emissions generated by local training and upload only the estimated carbon emission values to the server, without transmitting the carbon intensity or energy consumption associated with local training. When the server starts to carbon-aware select the clients for the upcoming training round \(t\) (\(t \geq T_{prep}\)), client \(i\) will receive a request from the server to send \(C_i^t\), the estimated carbon emissions of local training on client \(i\) for the upcoming training round \(t\). The detailed process of estimating the carbon emissions of local training for each client is as follows:
\begin{itemize}
    \item 1) Client $i$ estimates its availability period for the upcoming FL training round ($t$) based on the user's habitual daily behavior, including recent device usage and charging patterns. Specifically, the client identifies periods of the day when the device is typically connected to a charger and the user is unlikely to be actively using other applications. This estimation is performed locally on each client device, without sharing the user's behavioral information with any other entities in the network.
    \item 2) Client $i$ obtains $CI_i^t$, the average carbon intensity over its estimated availability period in the upcoming training round $t$, using hourly carbon intensity forecasts. Such forecasts can be obtained from existing carbon intensity forecasting services, for example, \cite{electricitymaps}, which provide hourly forecasts for different geographical regions. The methodology for forecasting carbon intensity is beyond the scope of this work.
    \item 3) Client \(i\) estimates \(E_i\), the energy consumption of its local training,  using the electric charge consumed ($Q_i$), which is recorded at the first time of local training of client \(i\), and the device voltage \(V_i\) according to \eqref{eq: electricity consumption calculation}. Since the local training dataset of each client is typically assumed to remain unchanged throughout the whole process of FL \cite{fedavg}, the electric charge consumed by local training on the same device can be assumed to be identical across different training rounds. Hence, \(Q_i\) and \(V_i\) are only recorded once in CA-FCS, as long as the size of the local training dataset remains unchanged.
    \item 4) Client \(i\) estimates \(C_i^t\) using \(E_i\) and \(CI_i^t\) according to \eqref{eq: carbon emissions calculation}.
\end{itemize}

After computing $C_i^t$, client \(i\) sends \(C_i^t\) to the server for the carbon-aware client selection process. Since \textit{the server receives only the estimated total carbon emissions} associated with each client's local training, rather than the carbon intensity values, it cannot directly infer the client's carbon intensity, device type, or geographical location from the reported information. As a result, CA-FCS mitigates privacy concerns associated with sharing clients’ sensitive information.

Unlike existing carbon-aware client selection strategies, \textit{CA-FCS comprehensively accounts for heterogeneity in both local training energy consumption and the carbon intensity across clients}, enabling more accurate estimation of carbon emissions. The heterogeneity in energy consumption arising from differences in client device hardware specifications is accounted for by using the recorded electric charge consumed during local training and the device voltage. These values vary across mobile devices, reflecting differences in their energy consumption characteristics. CA-FCS excludes the energy consumed by background applications and processes on mobile devices, ensuring that only the energy consumed during local training is considered when estimating each client's energy consumption. This approach improves the accuracy of local training energy consumption estimation compared to methods that rely on changes in the battery percentage reported by the device's operating system. Given the spatial and temporal variations in carbon intensity, and the fact that mobile devices typically perform local training only when connected to a charger and a stable Wi-Fi network \cite{greenfl}, CA-FCS considers the hourly carbon intensity at each client's location during its estimated availability period for local training, rather than relying on daily or annual average carbon intensity values. For example, if a client located in NSW, Australia, is available for local training from 3 a.m. to 6 a.m., the hourly carbon intensity values in NSW at 3 a.m., 4 a.m., and 5 a.m. are averaged to estimate the carbon intensity associated with the client's local training. This approach accounts for the spatial and temporal heterogeneity of carbon intensity while considering the time constraints on clients' participation in local training. Consequently, by combining more accurate estimates of local training energy consumption with fine-grained carbon intensity data, CA-FCS can achieve more precise carbon emissions estimation.

\section{Carbon-aware Fair Client Selection Design}
Carbon-aware client selection is designed to reduce carbon emissions of FL from client-side computing by prioritizing clients with lower carbon emissions associated with local training. However, consistently prioritizing clients with lower carbon emissions may lead to the exclusion of other clients, thereby reducing data diversity and degrading global model performance, particularly in non-IID data settings. To achieve sustainable and equitable FL, balancing environmental efficiency and fairness in client selection is essential. To address this challenge, CA-FCS introduces two constraints to prevent the repeated selection of certain clients and promote fairness in client participation: 1) selection frequency limits, which define the minimum and maximum number of times each client can be selected for local training; and 2) non-consecutive selection, which prevents the same client from being selected in two consecutive training rounds. The server specifies the number of clients \(N\), clients selected per round \(x\), and the minimum and maximum number of times each client can be selected, denoted by $K_{\min}$ and $K_{\max}$, respectively ($0 \le K_{min} \le K_{max}$). In the initial training rounds of CA-FCS, in each round, the server selects \(x\) clients from the total number of \(N\) clients sequentially to ensure that each client can participate in FL at least \(K_{min}\) times. The number of these initial training rounds \(T_{prep}\) is calculated by:
\begin{equation}
    T_{prep} = \lceil \frac{N}{x} \rceil \cdot K_{min}.
    \label{eq: preparation rounds calculation}
\end{equation}
When $K_{min} \ge 1$, in the initial training rounds, CA-FCS sequentially selects \(x\) clients per round to ensure that each client participates in training at least \(K_{\min}\) times. Specifically, at each initial training round $t \in [0,T_{prep}-1]$, clients with indices in the following cyclic interval are selected for training:
\begin{equation}
    S_t = [((t)x)\bmod N,\;((t+1)x)\bmod N],
    \label{eq:ClientsInitialRounds}
\end{equation}
where the modulo operation enables the selection process to wrap around to the first client after reaching the last client. When the last initial training round ($t=T_{prep}-1$) is finished, all $N$ clients have been selected to participate in FL training at least $K_{min}$ times. At round $t=T_{prep}-1$, if fewer than $x$ clients remain with selection counts below \(K_{min}\), the remaining slots are filled randomly from clients that have already reached \(K_{min}\) selected times. For instance, when $N=100$, $x=9$, and $K_{min}=1$, the index for each client ranges from 0 to 99 and $T_{prep}=12$. At the twelfth initial training round ($t=11$), only client 99 has a selection count less than $K_{min}=1$. Then the server will randomly select 8 clients from clients 0 to 98 to meet the required minimum number of selected clients per training round. By using this strategy, CA-FCS prevents certain clients from being permanently excluded, which may reduce the accuracy of the aggregated global model. Algorithm.\ref{alg:alg1} presents the process of initial rounds in CA-FCS.

During their first local-training participation, clients record consumed charge \(Q_i\) and voltage \(V_i\). These values are used to estimate the energy consumption of local training $E_i$ for selecting clients after the initial rounds. At round $t$ (\(t \geq T_{prep}\)), CA-FCS performs carbon-aware client selection under fairness constraints. To perform carbon-aware client selection, the server requires each client to estimate its carbon emissions of local training, \(C_i^t\), using \(E_i\) and \(CI_i^t\) according to \eqref{eq: carbon emissions calculation} and the procedure described in Section \ref{sec: process of estimating carbon emissions in CA-FCS}. Then, each client $i$ sends \(C_i^t\) to the server, which ranks all the clients in ascending order of their estimated carbon emissions. The server iteratively selects clients with minimum estimated carbon emissions, subject to two constraints: (1) the client was not selected in the previous training round, and (2) the client's selection count \(K_i\) is less than \(K_{\max}\). This process continues until \(x\) eligible clients have been selected. Algorithm.\ref{alg:alg2} presents the process after initial rounds in CA-FCS.

\begin{algorithm}
    \caption{The Initial Rounds in CA-FCS ($0 \leq t < T_{prep}$).}\label{alg:alg1}
    \begin{algorithmic}[1]
        \State Input: $S$, $K_{min}$, $N$, $x$
        \State $T_{prep} = \lceil \frac{N}{x} \rceil \cdot K_{min}$ 
        \For{$0 \leq t < T_{prep}$}
                \State  $S_t = [((t)x)\bmod N,\;((t+1)x)\bmod N]$
                \For {$i$ in $S_t$} 
                    \If{this is the first time client $i$ be selected}
                        \State Client $i$ records $Q_i$ and $V_i$.
                    \EndIf
                \EndFor
        \EndFor
\end{algorithmic}
\end{algorithm}

\begin{algorithm}
    \caption{The Rounds after Initial Rounds in CA-FCS ($t \geq T_{prep}$).}\label{alg:alg2}
    \begin{algorithmic}[1]
        \State Input: $S$, $K_{max}$, $N$, $x$
        \For {$t \geq T_{prep}$}
            \For{$i$ in $S$}
                    \State Client $i$ obtains \(CI_i^t\) according to its habitual.
                    \State $E_i = Q_i \cdot V_i$
                    \State $C_i^t = E_i \cdot CI_i^t$
                    \State Client $i$ sends $C_i^t$ to the server. 
                \EndFor
                \State The server sort $S$ in ascending order based on $C_i^t$
                \For{$i$ in sorted $S$}
                    \If{the server has selected $x$ clients}
                        \State Break
                    \ElsIf{$K_i \leq K_{max}$} 
                        \If{client $i$ was not selected in $t-1$}
                            \State Client $i$ is selected by the server.
                            \State $K_i = K_i + 1$
                        \EndIf
                    \EndIf
                \EndFor
        \EndFor
\end{algorithmic}
\end{algorithm}

\section{Experiments}
In this section, we explain the setup of our experiments in detail to evaluate the performance of CA-FCS. We examine the performance of CA-FCS and three other state-of-the-art client selection strategies by using 5 different datasets in both IID and non-IID data settings. The values of electricity-related information (electric charge consumed and voltage) of clients in our experiments are obtained from 3 different models of Google Pixel phones. 

\subsection{Experiment Settings}
\textit{Datasets}: The experiments are conducted on five image classification datasets with varying levels of complexity--MNIST\cite{mnist}, FMNIST\cite{fmnist}, CIFAR-10\cite{cifar}, CIFAR-100, and CINIC-10\cite{cinic10}. MNIST and FMNIST contain 28×28 grayscale images with 60,000 training and 10,000 test samples. CIFAR-10 consists of 60,000 color images belonging to 10 classes, with each image having a resolution of 32×32 pixels. CIFAR-100 follows the same image format but contains 100 classes instead of 10, with 600 images per class. CINIC-10 combines images from CIFAR-10 and ImageNet \cite{imagenet}. It contains 270,000 color images of size 32×32, evenly distributed across 10 classes. These datasets are used as the benchmark for testing performance in many FL-related papers.

\textit{Model architectures}: For MNIST and FMNIST, we employ a multi-layer perceptron (MLP) consisting of two hidden layers with 200 neurons each and ReLU activation functions (Table. \ref{tab:full connected model}). For CIFAR-10, CIFAR-100, and CINIC-10, we employ a convolutional neural network (CNN). The CNN architecture consists of two 5×5 convolutional layers with 32 and 64 output channels, respectively, each followed by a 2×2 max-pooling layer. The convolutional layers are followed by a fully connected layer with 512 ReLU-activated neurons and a softmax output layer for classification (Table. \ref{tab:CNN model}) \cite{fedavg}.
\begin{table}[b]
\caption{The model architecture of the fully connected model used for MNIST and FMNIST datasets}
\begin{tabular}{|c|c|c|}
\hline
Layer & Number of Neurons & Activation \\
\hline
Input & None & None \\
\hline
Dense & 200 & Relu \\
\hline
Dense & 200 & Relu \\
\hline
Dense & 10 & Softmax\\
\hline
\end{tabular}
\label{tab:full connected model}
\centering
\end{table}

\begin{table}[b]
    \caption{The model architecture of the CNN model used for CIFAR-10, CIFAR-100, and CINIC-10 datasets}
    \centering
    \begin{tabular}{|c|c|c|c|c|}
        \hline
        Layer & \makecell{Number of \\ Kernels} & \makecell{Size of \\ Kernel} & \makecell{Number of \\ Neurons} & Activation \\
        \hline
        Conv2D & 32 & (5,5) & None & Relu \\
        \hline
        Conv2D & 64 & (5,5) & None & Relu \\
        \hline
        Dense & None & None & 512 & Relu \\
        \hline
        Dense & None & None & 10 & Softmax \\
        \hline
    \end{tabular}
    \label{tab:CNN model}
\end{table}

\textit{Data partition}: The experiments consider both IID and non-IID data settings. Each dataset is equally partitioned into 100 subsets, where each subset serves as the local dataset of one client. Consequently, each client possesses the same number of samples: 600 for MNIST and FMNIST, 500 for CIFAR-10 and CIFAR-100, and 900 for CINIC-10. Two levels of non-IID data distributions are considered.
\begin{itemize}
    \item 1) Extreme non-IID setting: For MNIST, FMNIST, CIFAR-10, and CINIC-10, each client contains data from only two classes. For CIFAR-100, each client contains data from 20 classes.
    \item 2) Moderate non-IID setting: For MNIST, FMNIST, and CINIC-10, each client contains data from six classes. For CIFAR-10, each client contains data from five classes, while CIFAR-100 clients contain data from 50 classes.
\end{itemize}

\textit{Experiment assumptions}: To consider hardware heterogeneity across clients, three generations of Google Pixel phones (Google Pixel 5, Google Pixel 7a, and Google Pixel 8a) are used in our experiments \cite{fedtctf}. Google Pixel 6a is excluded due to reported thermal throttling issues \cite{pixel6a_recall}. These issues make it challenging to accurately record the energy consumption of local training. We assume that the clients using each type of smartphone account for equal proportions. On each phone, we apply a single model training procedure with the same dataset size and training configuration, then use Android Debug Bridge (ADB) command \cite{adb} to record the electric charge consumed by model training and device voltage on each smartphone. More details on how to use ADB commands are introduced in Section \ref{sec: carbon emissions from real devices}. 

Since carbon intensity varies substantially between daytime and nighttime periods, the time at which clients participate in FL training significantly affects the carbon emissions associated with local computation. In practical FL deployments, participating devices are typically required to be connected to both Wi-Fi and a power source to avoid unexpected dropouts \cite{greenfl}. Motivated by real-world user behavior patterns \cite{mammen2021wisleep}, we assume that 30\% of clients are available to start and finish local training during a randomly selected hour between 7 a.m. and 8 p.m., while the remaining 70\% are available during a randomly selected hour between 9 p.m. and 6 a.m. of the following day. To evaluate the robustness of the proposed method, additional experiments on FMNIST are conducted under alternative participation ratios of 10\%-90\%, 20\%-80\%, and 40\%-60\%. We further assume that each FL training round is completed within a single day. Time-varying carbon intensity values are obtained from Electricity Maps \cite{electricitymaps} and used to estimate carbon emissions. In this study, only carbon emissions associated with model training are considered.

\textit{Baseline and metric}: To evaluate the effectiveness of the proposed method, we compare CA-FCS against three baseline client selection strategies: Random Client Selection (Random CS)\cite{fedavg}, Carbon-Efficient FL (CEFL) \cite{cefl}, and Energy-Aware FL (EAFL) \cite{eafl}. Random CS uniformly samples an equal number of clients in each training round. CEFL prioritizes clients with lower carbon intensity and higher local training loss, while dynamically adjusting the number of selected clients according to the deviation of the accuracy curve. EAFL extends Oort \cite{oort} by incorporating both local training loss and energy consumption into the client utility metric. In the experiments using EAFL, the battery level of each client device was set to 100\% prior to training. This setting is based on our experimental assumption that all mobile devices must be connected to a power source and Wi-Fi network to participate in FL model training. In addition to Random CS, which is the most widely adopted FL client selection method, we include two additional baseline methods, CEFL and EAFL, as, \textit{to the best of our knowledge, they are the only existing methods that can be applied to minimize carbon emissions as an optimization objective with fine-grained carbon intensity.} The evaluation focuses on: 1) the highest achieved test accuracy, and 2) the carbon emissions generated by client-side computation within a fixed number of training rounds. The fixed training budget is set to 100 training rounds for MNIST, FMNIST, and CIFAR-10, and 200 training rounds for CIFAR-100 and CINIC-10.

For fairness evaluation, we compare the distribution of client selection frequencies across CA-FCS, Random CS, and fully carbon-aware client selection (CA-CS). CA-CS is a client selection strategy proposed by us for comparison that selects clients solely based on their carbon emissions from client-side computation. It removes all the constraints for fairness in CA-FCS and directly selects the clients with the lowest estimated carbon emissions produced by local training for each FL training round. The reason for proposing CA-CS and comparing it with Random CS and CA-FCS is to demonstrate the importance of incorporating fairness into carbon-aware client selection.

\textit{Other details}: We use Pydroid and PyTorch to code and train the model on Google Pixel phones. Table. \ref{tab:Configuration Parameters of the Experiments} summarizes the implementation details, including learning rate, batch size, and number of epochs.
\begin{table}
    \centering
    \caption{Configuration Parameters of the Experiments}
    \begin{tabular}{|c|c|}
        \hline
        Learning Rate for MNIST and FMNIST & 0.1 \\
        \hline
        Learning Rate for CIFAR-10, CIFAR-100 and CINIC-10 & 0.01 \\
        \hline
        Epochs & 5\\
        \hline
        Batch Size & 10 \\
        \hline
        Number of Total Clients ($N$) & 100 \\
        \hline
        Number of Selected Clients in Each Round ($x$) & 10 \\
        \hline
        $K_{min}$ & 1 \\
        \hline
        $K_{max}$ & 20 \\
        \hline
        Number of Training Rounds for MNIST, FMNIST and CIFAR-10 & 100 \\
        \hline
        Number of Training Rounds for CIAFR-100 and CINIC-10 & 200 \\
        \hline
        Types of Client Devices & 3\\
        \hline
    \end{tabular}
    \label{tab:Configuration Parameters of the Experiments}
\end{table}

\subsection{Estimating Energy Consumption from Google Pixel Phones Using ADB}
\label{sec: carbon emissions from real devices}
\begin{figure}
\centering
\includegraphics[width=72 mm, height=48 mm]{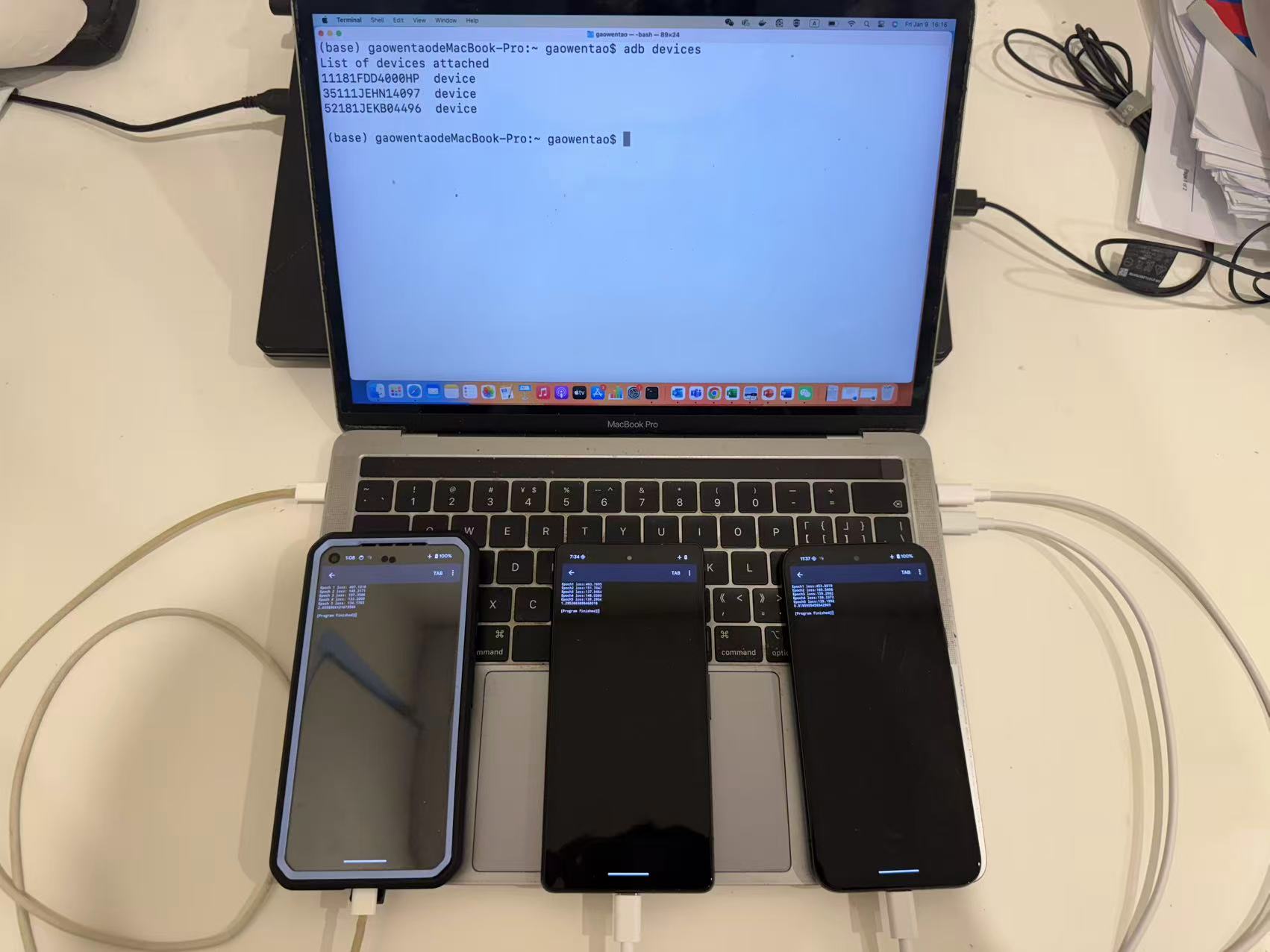}
\caption{Three Google Pixel phones connect to the computer with ADB to get the file--batterystats.txt, which
records the electric charge consumed for local training and device voltage.}
\label{fig: Devices Used for Measuring the Carbon Emissions from Edge Devices Local Training}
\end{figure}

\begin{figure}
    \centering
     \includegraphics[width=75 mm, height=30 mm]{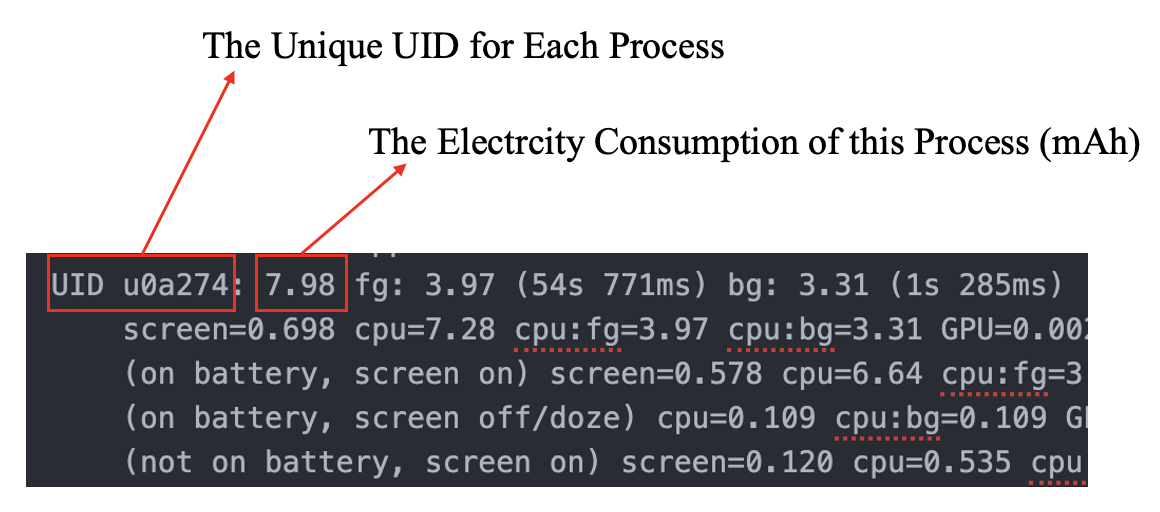}
    \caption{The contents of batterystats.txt, which is generated using ADB commands. It records the electric charge consumed by each operating process. Each process is associated with a unique UID, which can be used to identify the electric charge consumed by the process executing local training.}
    \label{fig:battery stats}
\end{figure}

The voltage of Google Pixel 5 and Google Pixel 8a is 3.87 $V$. The voltage of Google Pixel 7a is 3.85 $V$ (\(V_i\) used in our experiments). The electric charge consumed by Google Pixel 5, Google Pixel 7a, and Google Pixel 8a for local training (\(Q_i\) used in our experiments) is shown in Table. \ref{tab:device srtting}. Both the voltage and the electric charge consumed can be obtained by using ADB commands \cite{adb}. The specific steps to measure the electric charge consumed by a Google Pixel phone for local training are as follows:
\begin{itemize}
    \item 1) Enable developer options and USB debugging on the Google phones. And connect the Google phones to a computer that has ADB installed, such as in Fig. \ref{fig: Devices Used for Measuring the Carbon Emissions from Edge Devices Local Training}.
    \item 2) Reset battery stats before starting the local training with the command: 
    \begin{verbatim} adb shell dumpsys batterystats --reset\end{verbatim},
    to ensure that only the consumed charge of local training is recorded.
    \item 3) Disconnect the Google phones from the computer, and train the model with their local data.
    \item 4) Connect the Google phones to the computer again. And use the command:
    \begin{verbatim} adb shell dumpsys batterystats
    > batterystats.txt \end{verbatim}
    to obtain the report of battery stats.
    \item 5) Use the UID, which is a unique ID allocated for each process by the Android system, to find the consumed charge corresponding to local training, as shown in Fig.\ref{fig:battery stats}.
\end{itemize}

Then, the electric charge consumed and the device voltage obtained from these three smartphones are used as the electric charge consumed by local training and the device voltage of clients with the corresponding device type in our experiments. The energy consumption of each type of Google Pixel phone used in our experiments can be calculated using the electric charge consumed and the device voltage with \eqref{eq: electricity consumption calculation}. Table. \ref{tab:device srtting} shows the energy consumption of the Google Pixel phones with different datasets.

\begin{table*}
    \centering
    \caption{Electric Charge Consumed and Energy Consumption of Local Training by Using Different Google Pixel Phones with Various Datasets}
    \footnotesize\begin{tabular}{|c|c|c|c|c|c|c|}
    \hline
     & \multicolumn{2}{|c|}{Google Pixel 5} & \multicolumn{2}{|c|}{Google Pixel 7a} & \multicolumn{2}{|c|}{Google Pixel 8a} \\
    \hline
    Dataset & \makecell{Electric Charge \\ Consumed} & Energy Consumption & \makecell{Electric Charge \\ Consumed} & Energy Consumption & \makecell{Electric Charge \\ Consumed} & Energy Consumption \\ 
    \hline
    MNIST & 0.294 mAh & 0.00000114 kWh & 0.831 mAh & 0.00000320 kWh & 0.749 mAh & 0.00000300 kWh \\
    \hline
    FMNIST & 0.311 mAh & 0.00000120 kWh & 0.773 mAh & 0.00000298 kWh & 0.739 mAh & 0.00000286 kWh \\
    \hline
    CIFAR-10 & 1.68 mAh & 0.00000650 kWh & 3.76 mAh & 0.0000145 kWh & 5.04 mAh & 0.0000195 kWh \\
    \hline
    CIFAR-100 & 1.64 mAh & 0.00000635 kWh & 3.71 mAh & 0.0000144 kWh & 5.08 mAh & 0.0000197 kWh \\
    \hline
    CINIC-10 & 2.52 mAh & 0.00000975 kWh & 6.68 mAh & 0.0000257 kWh & 7.98 mAh & 0.0000309 kWh \\
    \hline
    \end{tabular}
    
    
    \label{tab:device srtting}
\end{table*}

\section{Evaluation}
This section evaluates CA-FCS from two perspectives. First, the highest achieved test accuracy and the client-side carbon emissions of CA-FCS are compared with those of Random CS, CEFL, and EAFL across five datasets and different data distributions. Second, the fairness of client selection in CA-FCS is assessed by comparing the selection frequencies of each client with those of Random CS, CA-FCS, and CA-CS.

\subsection{Accuracy and Carbon Emissions Comparison}
\begin{figure*}
    \centering
    \subfloat[MNIST IID]{
    \centering
        \includegraphics[width=57 mm, height=40 mm]{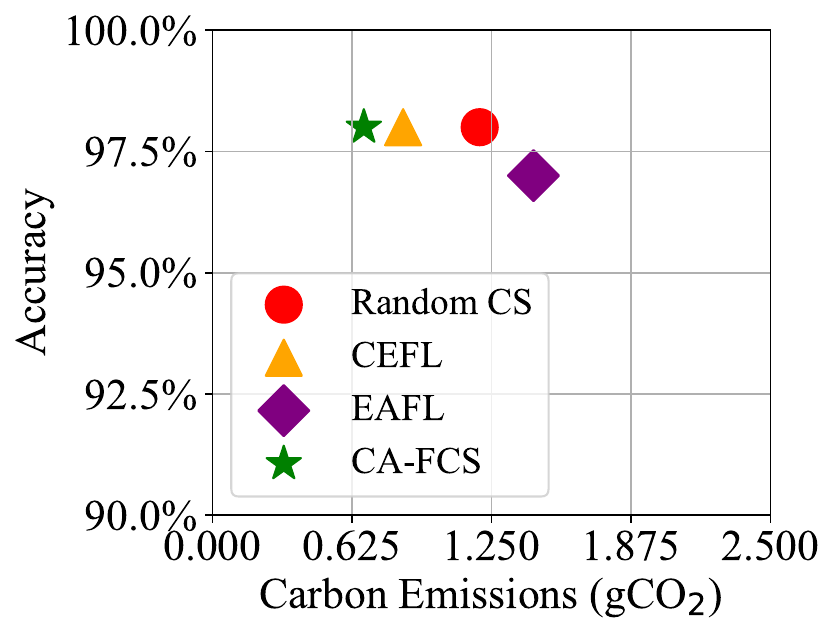}
        \label{fig: MNIST IID}
    }
    \subfloat[MNIST Non-IID (2 labels)]{
    \centering
        \includegraphics[width=57 mm, height=40 mm]{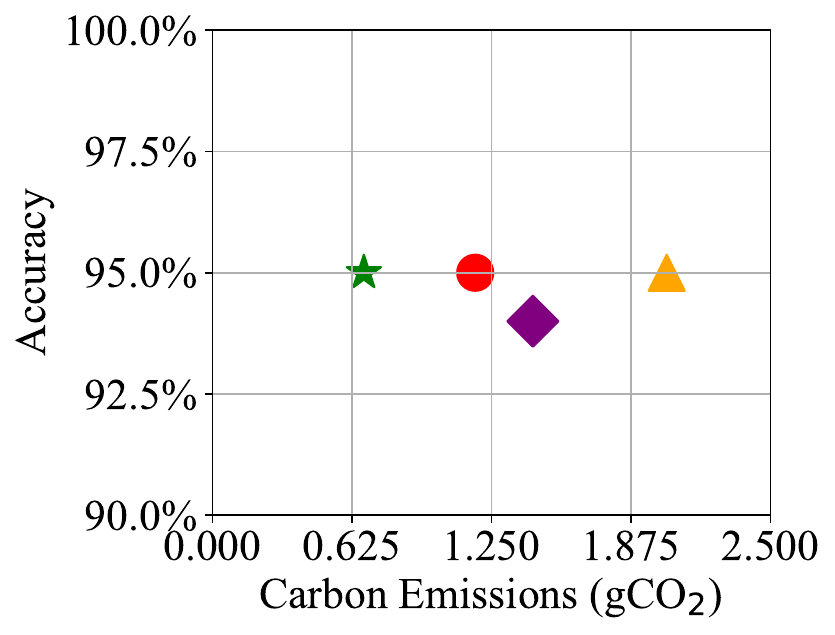}
        \label{fig: MNIST Non-IID (2 labels)}
    }
    \subfloat[MNIST Non-IID (6 labels)]{
    \centering
        \includegraphics[width=57 mm, height=40 mm]{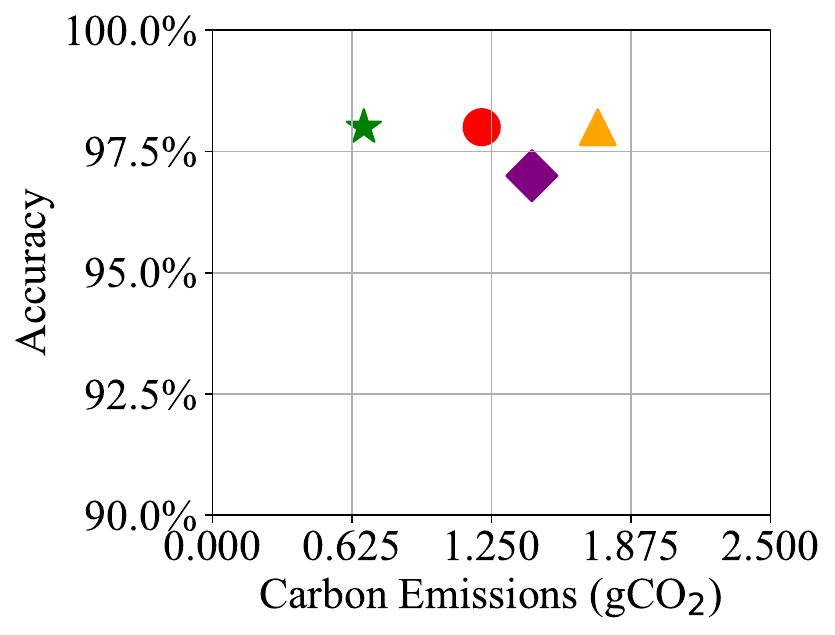}
        \label{fig: MNIST Non-IID (6 labels)}
    }
    \hfill
    \subfloat[FMNIST IID]{
    \centering
        \includegraphics[width=57 mm, height=40 mm]{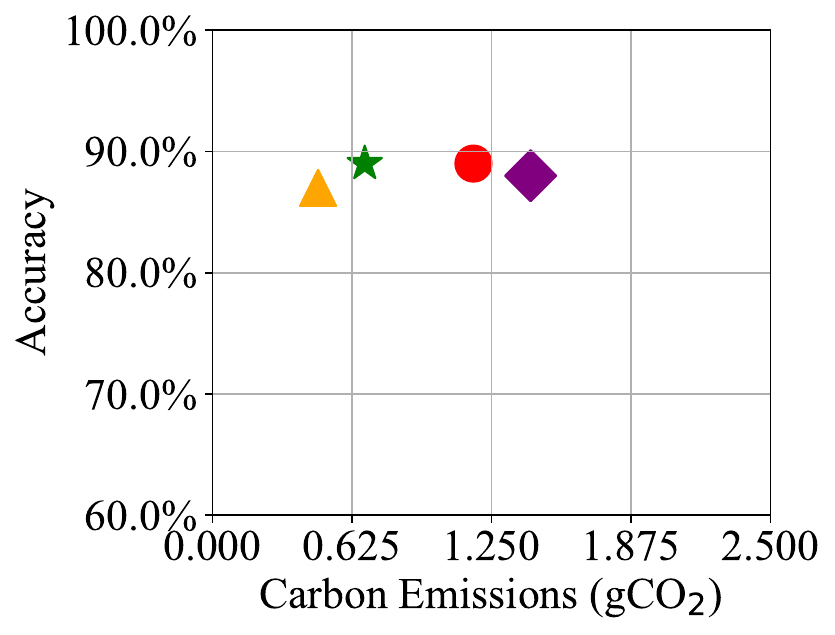}
        \label{fig: FMNIST IID}
    }
    \subfloat[FMNIST Non-IID (2 labels)]{
    \centering
        \includegraphics[width=57 mm, height=40 mm]{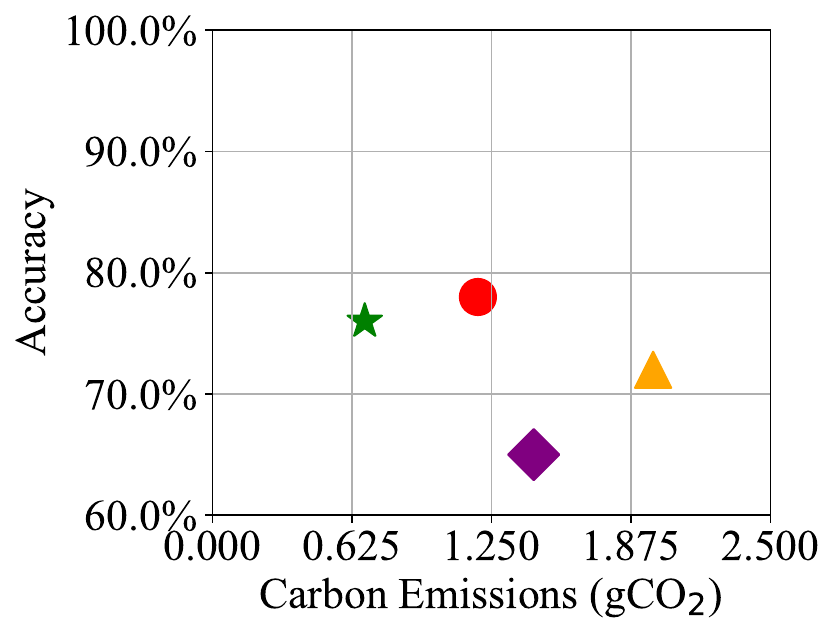}
        \label{fig: FMNIST Non-IID (2 labels)}
    }
    \subfloat[FMNIST Non-IID (6 labels)]{
    \centering
        \includegraphics[width=57 mm, height=40 mm]{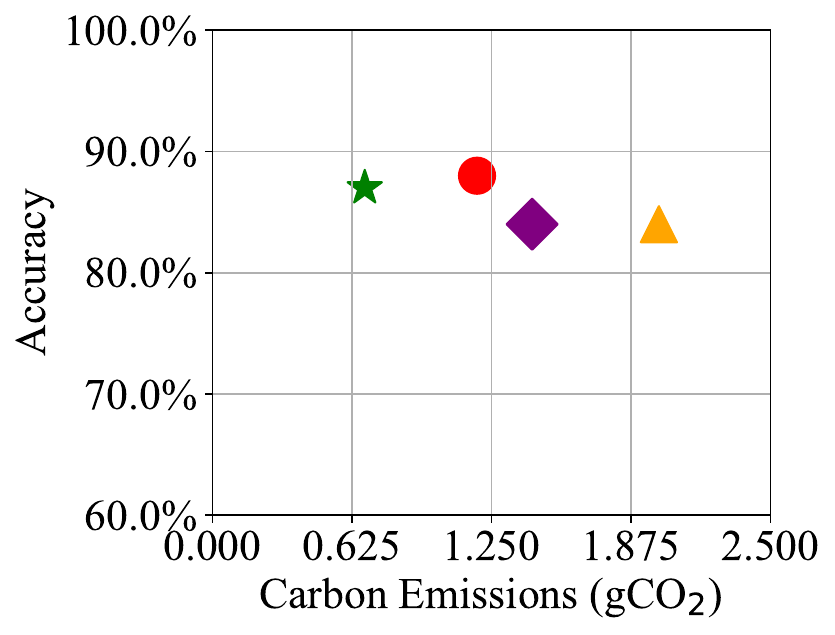}
        \label{fig: FMNIST Non-IID (6 labels)}
    }
    \caption{Highest achieved accuracy vs. carbon emissions from client-side computing for Random CS, CA-FCS, CEFL, and EAFL of 100 training rounds on MNIST and FMNIST. To the best of our knowledge, CEFL and EAFL are the only existing methods that can be applied to minimize carbon emissions as an optimization objective with fine-grained carbon intensity.}
    \label{fig: experimental results of MNIST and FMNNIST}
\end{figure*}

\textit{Performance on MNIST and FMNIST}: Fig. \ref{fig: experimental results of MNIST and FMNNIST} reports the highest test accuracy and client-side carbon emissions accumulated over 100 FL training rounds for Random CS, CEFL, EAFL, and CA-FCS with different data distributions on MNIST and FMNIST. On MNIST (Figs. \ref{fig: MNIST IID}-\ref{fig: MNIST Non-IID (6 labels)}), CA-FCS consistently incurs the lowest carbon emissions among all methods. Specifically, it reduces client-side carbon emissions by approximately 42\%-44\% relative to Random CS without sacrificing accuracy. Under IID, non-IID (2 labels), and non-IID (6 labels) settings, CA-FCS attains peak accuracies of 98\%, 95\%, and 98\%, respectively, which match those of Random CS and CEFL, and exceed those of EAFL. A similar trend is observed on FMNIST (Figs. \ref{fig: FMNIST IID}-\ref{fig: FMNIST Non-IID (6 labels)}). CA-FCS yields the lowest carbon emissions under both non-IID settings and the second-lowest emissions under the IID setting. Compared with Random CS, which achieves the highest accuracy in all data distributions, CA-FCS reduces carbon emissions by more than 41\% while maintaining competitive accuracy. Although CEFL attains slightly lower emissions than CA-FCS under the IID setting, its accuracy degrades noticeably under non-IID distributions. In contrast, CA-FCS preserves accuracy close to that of Random CS, with only a marginal reduction of approximately 1\%-2\% under non-IID settings.

The inferior accuracy of CEFL and EAFL is primarily attributed to insufficient fairness in client selection. Both methods repeatedly prioritize clients with low carbon intensity or low energy consumption, whereas clients associated with higher emissions remain excluded for extended periods. This selection bias reduces data diversity during training, which is particularly detrimental under non-IID distributions. The negative impact is further amplified when training on more complex datasets. The performance of CEFL depends strongly on its convergence behavior. CEFL adjusts the number of selected clients according to the slope of the accuracy curve. When convergence occurs early, fewer clients are selected in subsequent training rounds, while lowering the total emissions. However, under delayed convergence, particularly in non-IID settings, CEFL tends to select substantially more clients, and thereby incurs higher carbon emissions from local training. EAFL reduces energy consumption but does not consistently reduce carbon emissions relative to Random CS. Carbon emissions are jointly determined by energy consumption and carbon intensity. Optimizing only energy efficiency cannot guarantee carbon efficiency. 

Finally, both CEFL and EAFL require evaluating the local training loss of all clients before each selection round. The associated carbon emissions incurred by this evaluation process are excluded in our analysis. However, in large-scale FL deployments involving billions of edge devices, such additional computation introduces non-negligible environmental overhead.    

\begin{figure*}
    \centering
    \subfloat[CIFAR-10 IID]{
    \centering
        \includegraphics[width=57 mm, height=40 mm]{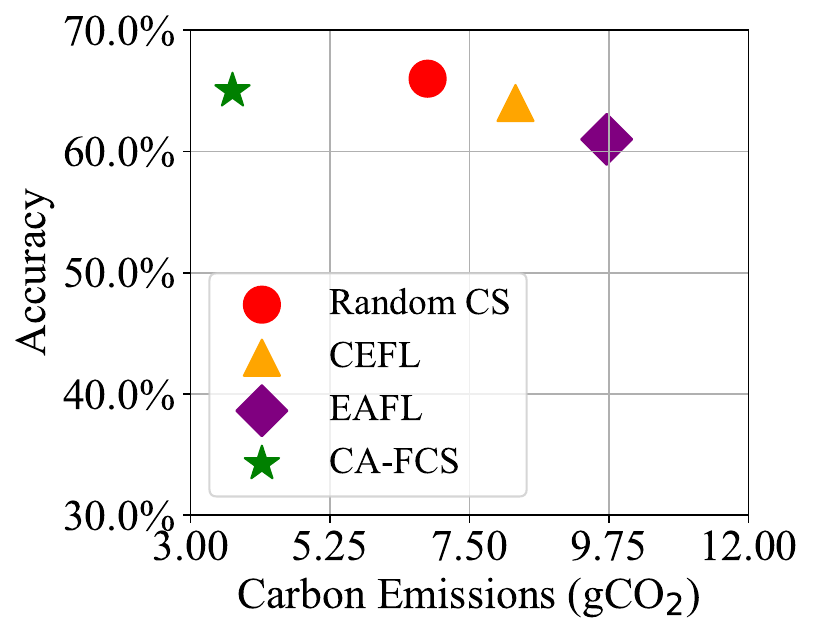}
        \label{fig: CIFAR10 IID}
    }
    \subfloat[CIFAR-10 Non-IID (2 labels)]{
    \centering
        \includegraphics[width=57 mm, height=40 mm]{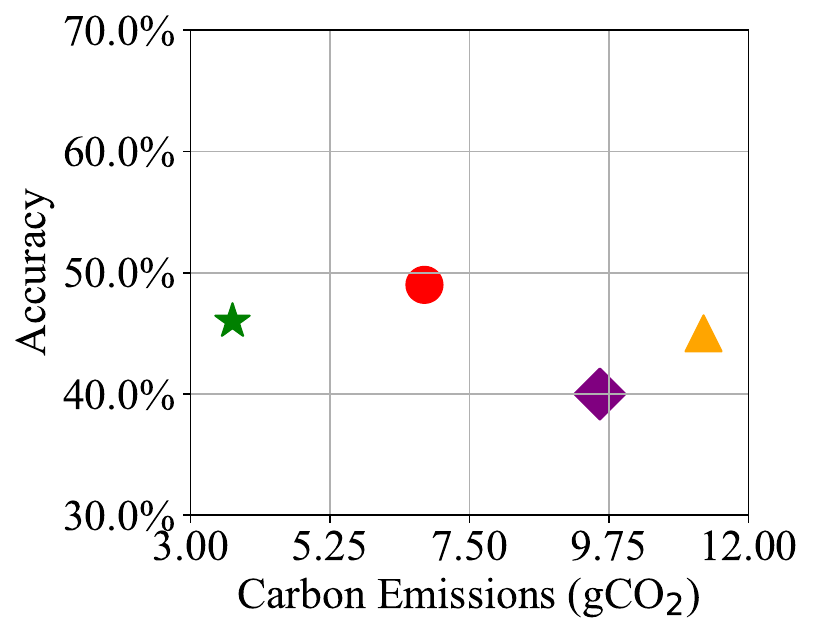}
        \label{fig: CIFAR10 Non-IID (2 labels)}
    }
    \subfloat[CIFAR-10 Non-IID (5 labels)]{
    \centering
        \includegraphics[width=57 mm, height=40 mm]{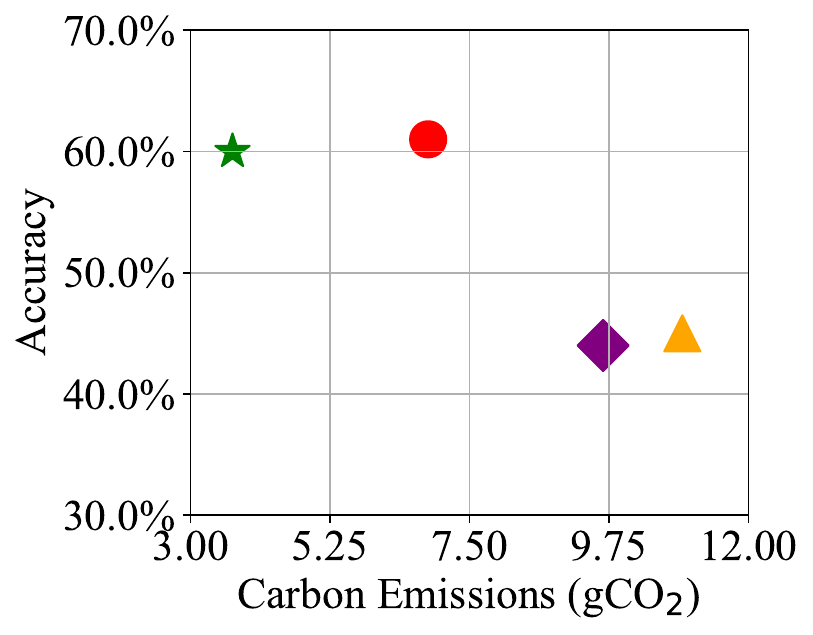}
        \label{fig: CIFAR10 Non-IID (5 labels)}
    }
    \hfill
    \subfloat[CIFAR-100 IID]{
    \centering
        \includegraphics[width=57 mm, height=40 mm]{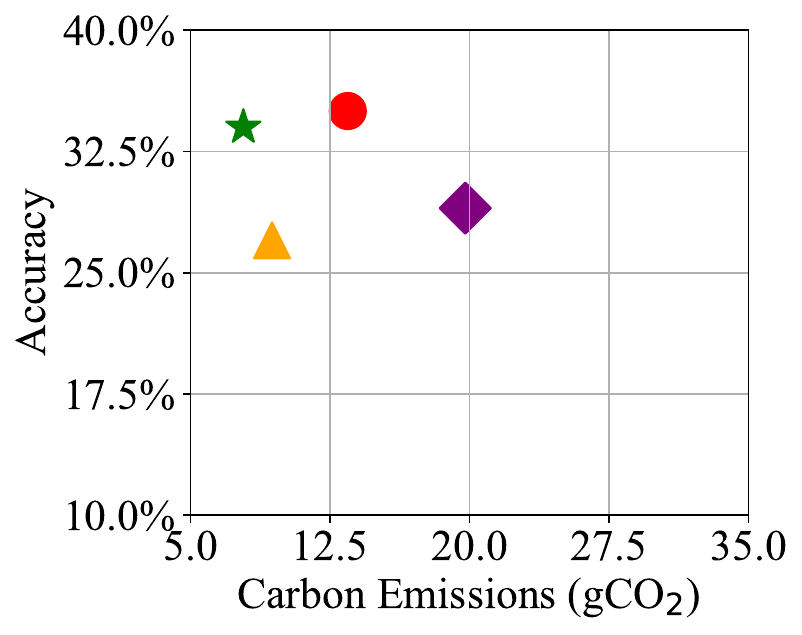}
        \label{fig: CIFAR100 IID}
    }
    \subfloat[CIFAR-100 Non-IID (20 labels)]{
    \centering
        \includegraphics[width=57 mm, height=40 mm]{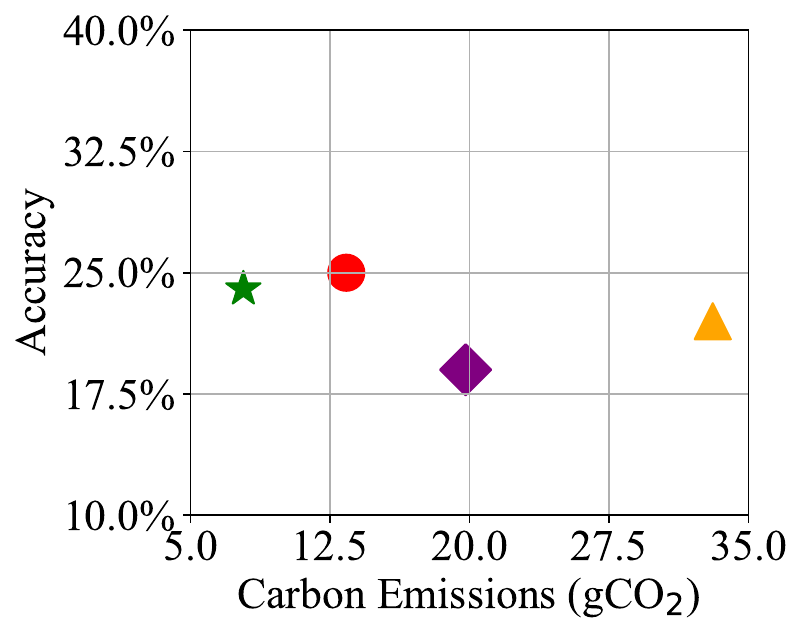}
        \label{fig: CIFAR100 Non-IID (20 labels)}
    }
    \subfloat[CIFAR-100 Non-IID (50 labels)]{
    \centering
        \includegraphics[width=57 mm, height=40 mm]{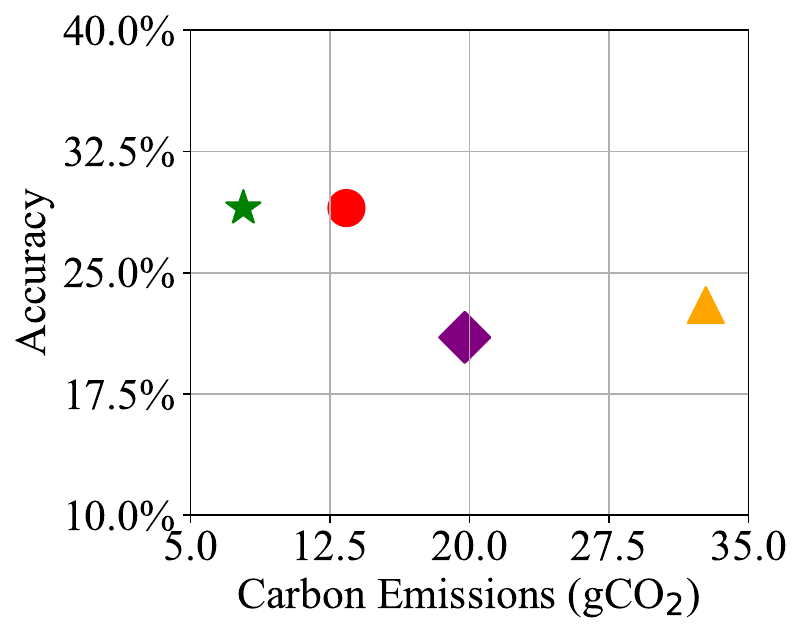}
        \label{fig: CIFAR100 Non-IID (50 labels)}
    }
    \hfill
    \subfloat[CINIC-10 IID]{
    \centering
        \includegraphics[width=57 mm, height=40 mm]{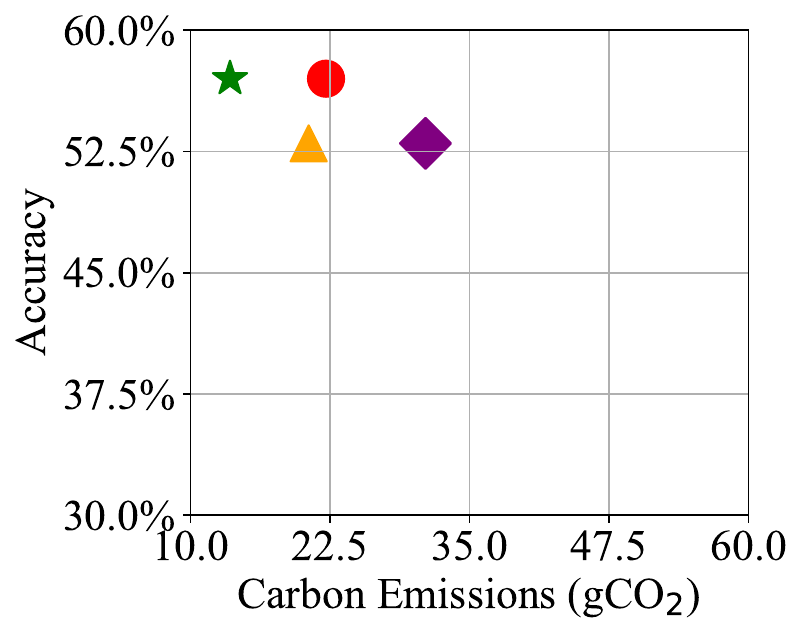}
        \label{fig: CINIC10 IID}
    }
    \subfloat[CINIC-10 Non-IID (2 labels)]{
    \centering
        \includegraphics[width=57 mm, height=40 mm]{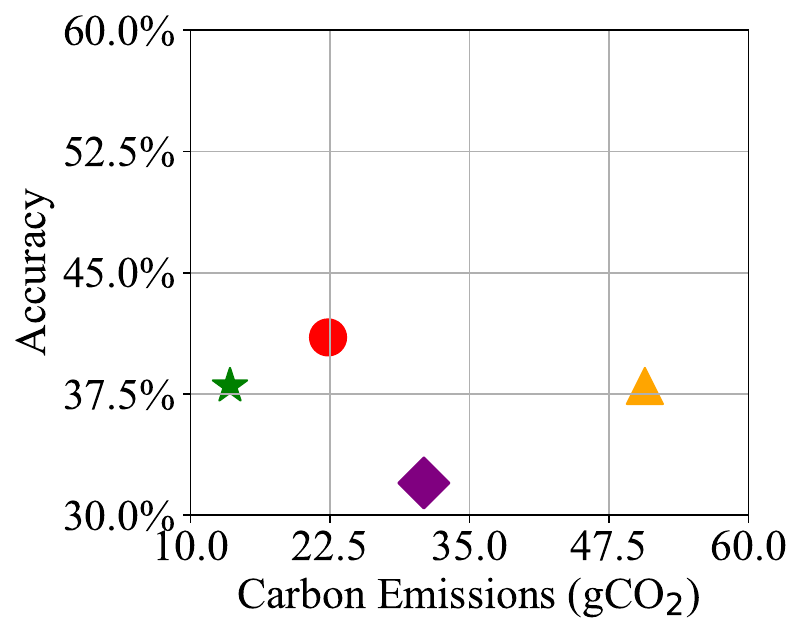}
        \label{fig: CINIC10 Non-IID (2 labels)}
    }
    \subfloat[CINIC-10 Non-IID (6 labels)]{
    \centering
        \includegraphics[width=57 mm, height=40 mm]{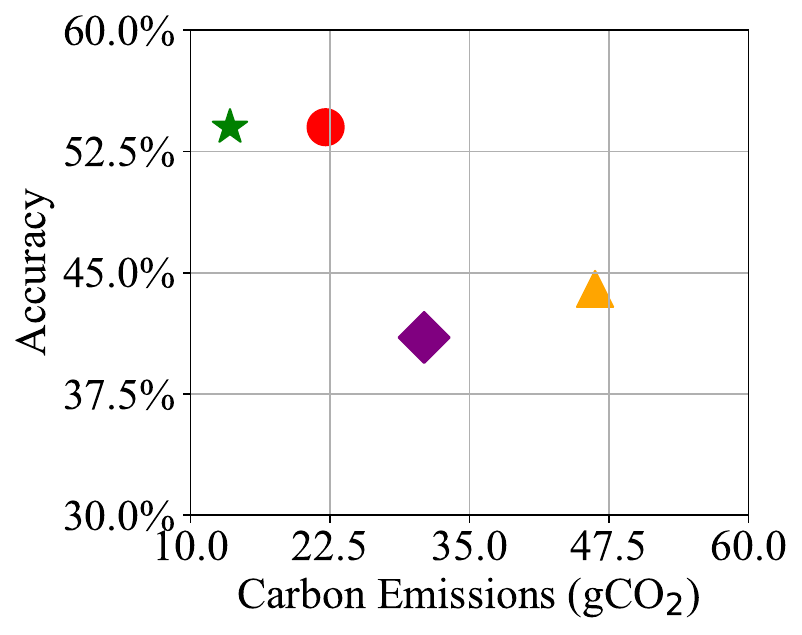}
        \label{fig: CINICR10 Non-IID (6 labels)}
    }
    \caption{Highest achieved accuracy vs. carbon emissions from client-side computing for Random CS, CA-FCS, CEFL, and EAFL of 100 training rounds on CIFAR-10, and 200 training rounds on CIFAR-100 and CINIC-10.}
    \label{fig: highest accuracy and carbon emissions in more complicated datasets}
\end{figure*}

\textit{Performance on CIFAR-10, CIFAR-100 and CINIC-10}: Fig. \ref{fig: highest accuracy and carbon emissions in more complicated datasets} presents the results on CIFAR-10, CIFAR-100, and CINIC-10, which are more complex datasets. On CIFAR-10, CA-FCS consistently achieves the lowest client-side carbon emissions across all data distributions, reducing them by approximately 46\% relative to Random CS. Meanwhile, its accuracy remains comparable to Random CS, with a maximum degradation of only 3\% under the extreme non-IID (2 labels) setting. On CIFAR-100, CA-FCS again yields the lowest carbon emissions among all methods. Compared with Random CS, which attains the highest accuracy overall, CA-FCS reduces client-side carbon emissions by 41\%-42\% at the cost of only approximately 1\% accuracy degradation under IID and non-IID (20 labels) settings. Similar observations are obtained on CINIC-10, where CA-FCS achieves the lowest carbon emissions and the second-best accuracy among all methods. With the non-IID (2 labels) setting, its highest accuracy is only 3\% lower than that of Random CS, while carbon emissions from client-side computation are reduced by approximately 39\%. These results indicate that the advantage of CA-FCS persists on more complex datasets, where the accuracy reduction of carbon-aware selection remains small relative to the carbon emissions savings.

\begin{figure}
    \centering
    \includegraphics[width=75 mm, height=42 mm]{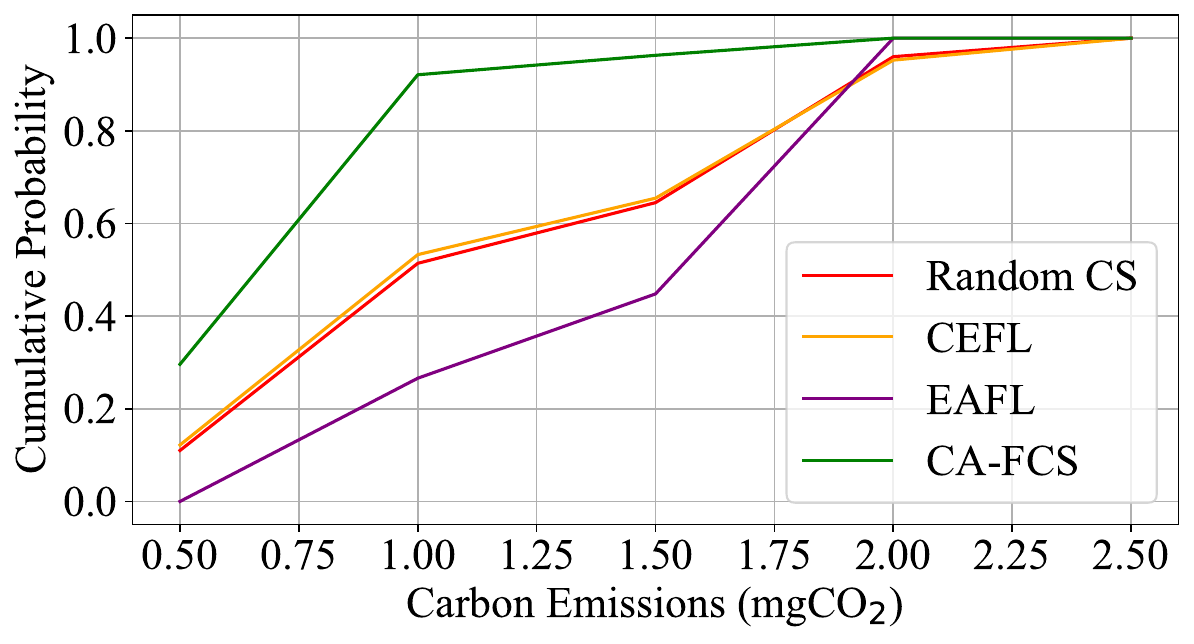}
    \caption{Cumulative distribution of the carbon emissions of selected clients under four client selection methods (Random CS, CEFL, EAFL, CA-FCS) on FMNIST over 100 rounds. Emissions are binned in intervals of 0.5 $mgCO_2$, so the cumulative probability increases in discrete steps at 0.5, 1, 1.5, 2, and 2.5 ($mgCO_2$). Curves that rise earlier indicate a stronger preference for low-emission clients.}
    \label{fig: CDF chart}
\end{figure}

\textit{Distribution of per-client carbon emissions}: Fig. \ref{fig: CDF chart} presents the cumulative distribution function (CDF) of the per-client carbon emissions of the clients selected by Random CS, CEFL, EAFL, and CA-FCS over 100 training rounds on FMNIST. For comparison, the emissions of the selected clients are grouped into six intervals: $\leq0.5$ $mg$, $(0.5,1]$ $mg$, $(1,1.5]$ $mg$, $(1.5,2]$ $mg$, $(2,2.5]$ $mg$, and $> 2.5$ $mgCO_2$. The cumulative probability is therefore evaluated only at the bin boundaries on the x-axis (0.5, 1, 1.5, 2, and 2.5 $mgCO_2$). The results show that CA-FCS prioritizes low-emissions clients substantially more effectively than the other methods. More than 80\% of the clients selected by CA-FCS generate no more than 1 $mgCO_2$ during local training. In comparison, fewer than 60\% of the clients that generate $CO_2$ less than or equal to 1$mg$ are selected by Random CS and CEFL, and the percentage of EAFL is much lower than 40\%. These observations confirm that CA-FCS more effectively concentrates training participation on low-emission clients while still preserving the fairness constraints.

\textit{Robustness under different temporal availability distributions}: Table. \ref{tab: different time distributions} reports the results on FMNIST under different temporal availability distributions. In addition to the default setting, in which 30\% of clients participate during daytime hours and 70\% during nighttime hours, we consider participation ratios of 10\%-90\%, 20\%-80\%, and 40\%-60\%. Across all evaluated settings, CA-FCS reduces client-side carbon emissions by at least 38\% relative to Random CS while limiting accuracy degradation to at most 3\%. Although CEFL attains lower carbon emissions from client-side computing with the IID distribution, its accuracy is lower. Moreover, its carbon emissions with non-IID distributions are substantially higher than those of CA-FCS. These results demonstrate the robustness of CA-FCS to heterogeneous temporal participation patterns.

\begin{table*}
    \centering
    \caption{Highest Accuracy and Carbon Emissions of 100 Training Rounds on FMNIST with Different Temporal Availability Distributions}
    \footnotesize\begin{tabular}{|c|c|c|c|c|c|c|}
    \hline
      \multicolumn{7}{|c|}{10\%(7 am to 8 pm)-90\%(9 pm to 6 am next day)} \\
    \hline
     & \multicolumn{3}{|c|}{Highest Accuracy in 100 Training Rounds} & \multicolumn{3}{|c|}{\makecell{Carbon Emissions from Client-Side Computing \\ of 100 Training Rounds ($gCO_2$)}} \\
    \hline
     Client Selection Methods & IID & Non-IID (2 labels) & Non-IID (6 labels) & IID & Non-IID (2 labels) & Non-IID (6 labels) \\
     \hline
     Random CS & 89\% & 78\% & 88\% & 1.201 & 1.182 & 1.155 \\
     \hline
    CEFL & 87\% & 74\% & 84\% & 0.591 & 2.027 & 1.768 \\
    \hline
    EAFL & 88\% & 67\% & 84\% & 1.416 & 1.427 & 1.431 \\
    \hline
    \textbf{CA-FCS} & \textbf{88\%} & \textbf{76\%} & \textbf{87\%} & \textbf{0.716} & \textbf{0.716} & \textbf{0.716} \\
    \hline
    \multicolumn{7}{|c|}{20\%(7 am to 8 pm)-80\%(9 pm to 6 am next day)} \\
    \hline
     & \multicolumn{3}{|c|}{Highest Accuracy in 100 Training Rounds} & \multicolumn{3}{|c|}{\makecell{Carbon Emissions from Client-Side Computing \\ of 100 Training Rounds ($gCO_2$)}} \\
    \hline
     Client Selection Methods & IID & Non-IID (2 labels) & Non-IID (6 labels) & IID & Non-IID (2 labels) & Non-IID (6 labels) \\
     \hline
     Random CS & 89\% & 78\% & 88\% & 1.174 & 1.138 & 1.171 \\
     \hline
    CEFL & 87\% & 68\% & 85\% & 0.448 & 1.772 & 1.809 \\
    \hline
    EAFL & 88\% & 72\% & 84\% & 1.411 & 1.424 & 1.421 \\
    \hline
    \textbf{CA-FCS} & \textbf{88}\% & \textbf{76\%} & \textbf{87\%} & \textbf{0.697} & \textbf{0.697} & \textbf{0.697} \\
    \hline
    \multicolumn{7}{|c|}{30\%(7 am to 8 pm)-70\%(9 pm to 6 am next day)} \\
    \hline
     & \multicolumn{3}{|c|}{Highest Accuracy in 100 Training Rounds} & \multicolumn{3}{|c|}{\makecell{Carbon Emissions from Client-Side Computing \\ of 100 Rounds ($gCO_2$)}} \\
    \hline
     Client Selection Methods & IID & Non-IID (2 labels) & Non-IID (6 labels) & IID & Non-IID (2 labels) & Non-IID (6 labels) \\
     \hline
     Random CS & 89\% & 78\% & 88\% & 1.169 & 1.189 & 1.185 \\
     \hline
    CEFL & 87\% & 72\% & 84\% & 0.473 & 1.974 & 2.000 \\
    \hline
    EAFL & 88\% & 65\% & 84\% & 1.425 & 1.439 & 1,432 \\
    \hline
    \textbf{CA-FCS} & \textbf{89\%} & \textbf{76\%} & \textbf{87\%} & \textbf{0.682} & \textbf{0.682} & \textbf{0.682} \\
    \hline
    \multicolumn{7}{|c|}{40\%(7 am to 8 pm)-60\%(9 pm to 6 am next day)} \\
    \hline
     & \multicolumn{3}{|c|}{Highest Accuracy in 100 Training Rounds} & \multicolumn{3}{|c|}{\makecell{Carbon Emissions from Client-Side Computing \\ of 100 Training Rounds ($gCO_2$)}} \\
    \hline
     Client Selection Methods & IID & Non-IID (2 labels) & Non-IID (6 labels) & IID & Non-IID (2 labels) & Non-IID (6 labels) \\
     \hline
     Random CS & 88\% & 79\% & 88\% & 1.181 & 1.150 & 1.169 \\
     \hline
    CEFL & 87\% & 70\% & 86\% & 0.514 & 1.934 & 1.925 \\
    \hline
    EAFL & 88\% & 70\% & 85\% & 1.427 & 1.430 & 1.436 \\
    \hline
    \textbf{CA-FCS} & \textbf{88\%} & \textbf{76\%} & \textbf{87\%} & \textbf{0.664} & \textbf{0.664} & \textbf{0.664} \\
    \hline
    \end{tabular}
    \label{tab: different time distributions}
\end{table*}

\textit{Summary}: CA-FCS effectively reduces carbon emissions from client-side computation across diverse datasets and data distributions while maintaining competitive accuracy, thereby supporting sustainable FL on resource-constrained mobile devices. Across all experiments, \textit{CA-FCS reduces carbon emissions from client-side computing by approximately 38\%-46\% relative to Random CS, with a decrease of at most 3\% in the highest achieved accuracy.}

\subsection{Fairness Evaluation}
Clients are assumed to participate in FL training at fixed time periods each day, and the temporal variation of carbon intensity within a given region generally follows a stable daily pattern. Moreover, each client’s local dataset remains unchanged throughout the FL process, which results in approximately constant energy consumption for local training. Consequently, the carbon emissions generated by the same client remain relatively consistent across training rounds.

By using CA-CS, client selection is determined solely by the estimated carbon emissions of local training. Clients with lower carbon emissions are therefore repeatedly prioritized, whereas clients with higher emissions are systematically excluded, which introduces substantial unfairness. In contrast, CA-FCS explicitly incorporates fairness constraints into client selection. Specifically, CA-FCS enforces both minimum and maximum participation frequencies for each client and prevents clients from being selected in consecutive training rounds. These constraints mitigate the over-selection of low-emission clients and ensure that high-emission clients are not permanently excluded from FL training.

\begin{figure}
    \centering
    \includegraphics[width=2.5 in]{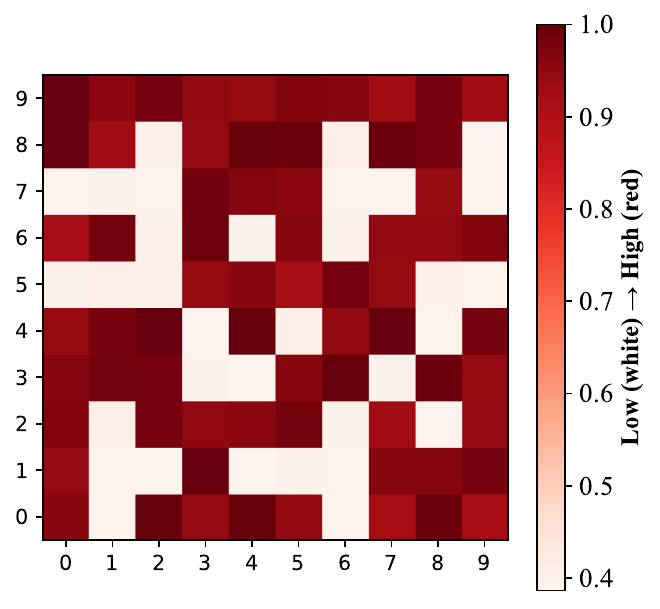}
    \caption{The carbon emissions level of each client when operating a local training of a fully connected model with 600 samples from FMNIST. Each small grid cell represents a client, and darker colors indicate higher carbon emissions generated during local training.}
    \label{fig: Carbon Emissions Grids}
\end{figure}

\begin{figure*}
    \centering
    \subfloat[Random CS]{
    \centering
        \includegraphics[width=51 mm, height=51 mm]{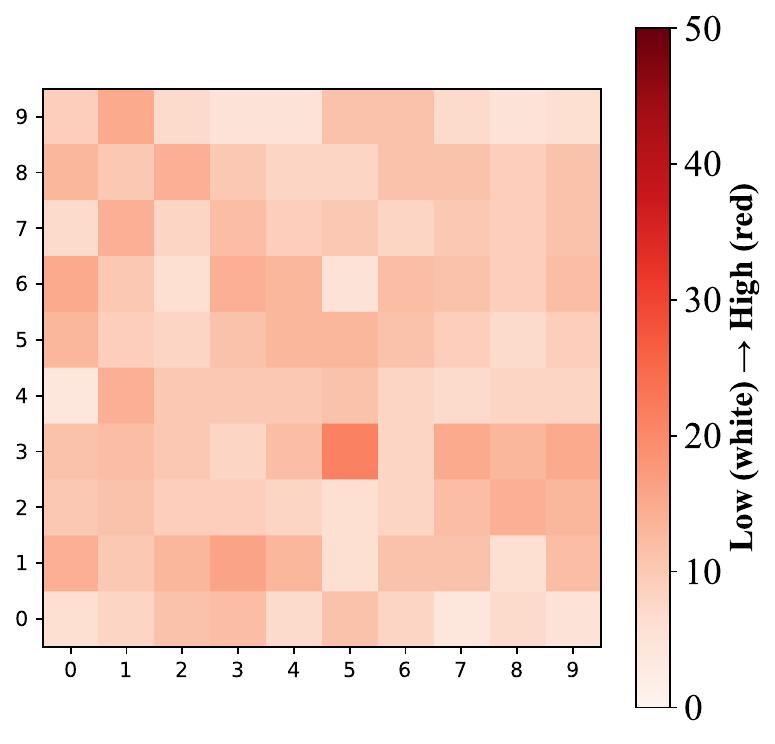}
        \label{fig: Random CS Grids}
    }
    \subfloat[CA-FCS]{
    \centering
        \includegraphics[width=51 mm, height=51 mm]{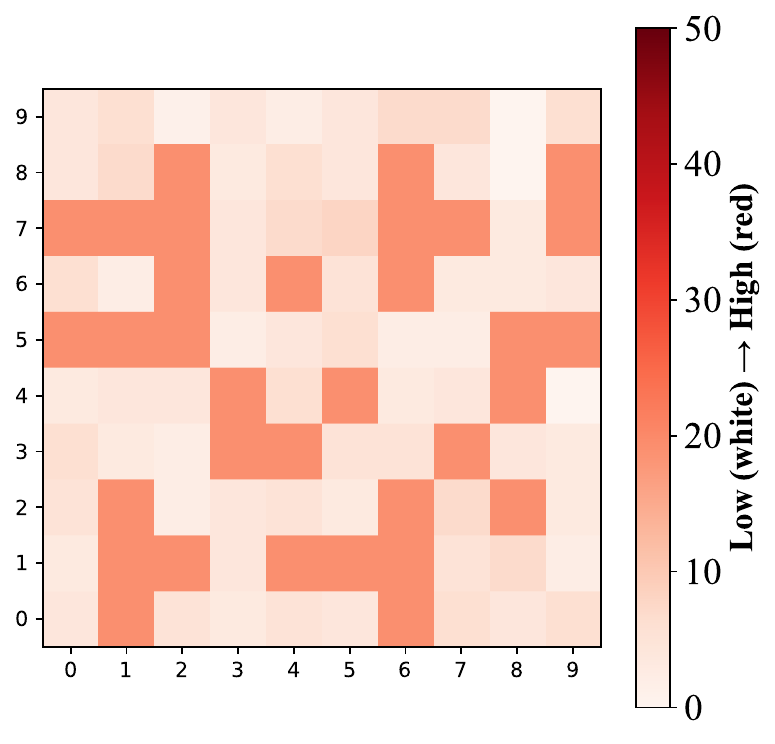}
        \label{fig: CA-FCS Grids}
    }
    \subfloat[CA-CS]{
    \centering
        \includegraphics[width=51 mm, height=51 mm]{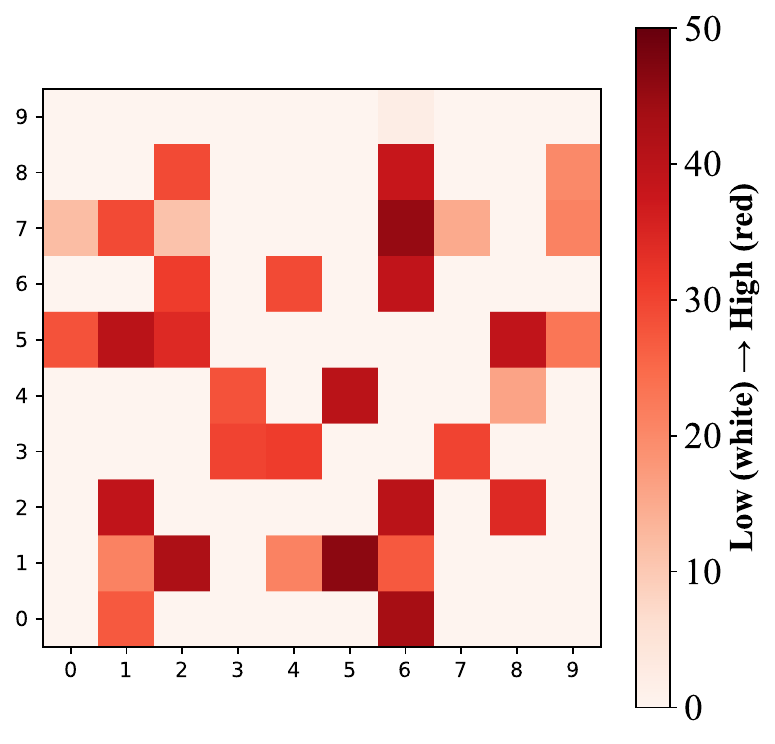}
        \label{fig: CA-CS Grids}
    }
    \caption{The participation times of each client in 100 training rounds of FL using different client selection strategies. Each small grid cell represents a client, and darker colors indicate a higher selection frequency.}
    \label{fig: participation frequency}
\end{figure*}

Fig. \ref{fig: Carbon Emissions Grids} shows the level of carbon emissions generated by each client operating a local training. Fig. \ref{fig: Random CS Grids}, Fig. \ref{fig: CA-FCS Grids}, and Fig. \ref{fig: CA-CS Grids} present the client participation frequencies over 100 FL training rounds under Random CS, CA-FCS, and CA-CS, respectively. Comparing Fig. \ref{fig: Carbon Emissions Grids} with Fig. \ref{fig: participation frequency}, it can be observed that Random CS maintains approximately uniform participation frequencies across all clients, regardless of their carbon emissions. In contrast, CA-CS strongly favors clients with lower emissions. Some low-emission clients are selected more than 40 times, while a substantial number of high-emission clients are never selected throughout the entire FL process. This demonstrates that purely carbon-driven selection can lead to severe participation imbalance. CA-FCS achieves a more balanced trade-off between carbon efficiency and fairness. As shown in Fig. \ref{fig: CA-FCS Grids}, clients with lower carbon emissions are still preferentially selected more frequently, enabling carbon reduction benefits, while clients with higher emissions are also guaranteed opportunities to participate in training. The maximum participation frequency under CA-FCS remains below 20 training rounds, and every client is selected at least once during the FL process.

\section{Discussion and Future Work} 
CA-FCS limits each client's participation times between \(K_{min}\) and \(K_{max}\) and prohibits selection in consecutive rounds. However, in this study, \(K_{min}\) and \(K_{max}\) are fixed to 1 and 20, respectively, regardless of the underlying data distribution. In future work, these fairness constraints could be adaptively adjusted according to data heterogeneity. Increasing the minimum numbers while reducing the maximum numbers generally leads to a fairer selection process, as it encourages more balanced participation among clients throughout FL training. Such adaptive mechanisms may be particularly beneficial under highly non-IID settings.

The experiments in this paper are conducted with real mobile devices and carbon intensity from the real world. However, this still differs from deployment in a fully realistic environment. For example, in practical FL systems, client selection may not be performed before every communication round. Instead, a single client selection process may be conducted, and the selected clients participate in FL for multiple subsequent rounds. Under such conditions, our method can be adapted into a periodic process for carbon-aware client selection. Specifically, all clients are first selected to participate in FL once and provide their carbon emissions from local training, after which the server selects clients based on the carbon emissions generated during their local training in this participation. The selected clients then participate in the subsequent several communication rounds of training. This constitutes a complete cycle, which is repeated multiple times until the training objective is achieved. The fairness constraints in CA-FCS would also be modified. The minimum participation requirement for each client can be removed, since the first step of each cycle—where all clients are selected to participate in FL once—already guarantees the participation frequency of all clients. Clients that have reached the maximum participation limit will no longer be selected by the carbon-aware client selection process in subsequent cycles. Furthermore, clients selected in the current cycle will not be selected again by the carbon-aware client selection process in the immediately following cycle. 

This work focuses primarily on reducing the carbon emissions generated by client-side computation through carbon-aware client selection. Other important sources of carbon emissions in FL systems, such as communication overhead between clients and the central server, are not considered in this study. Future research could therefore explore communication-aware carbon optimization strategies. Integrating both computation-side and communication-side optimization may enable the development of a more comprehensive and sustainable framework for Green FL.

\section{Conclusion}
In this paper, we propose CA-FCS, a fair carbon-aware client selection strategy for FL that aims to reduce the carbon emissions generated by client-side computation during local training. Unlike existing approaches that primarily rely on energy consumption, CA-FCS explicitly estimates the carbon emissions associated with local training and prioritizes clients with lower emissions during the selection process. To preserve fairness and maintain model performance, CA-FCS incorporates participation constraints that regulate the selection frequency of each client. These fairness mechanisms prevent the repeated selection of a small subset of low-emission clients and enable balanced participation across the FL process. Consequently, CA-FCS is able to maintain competitive model accuracy within a limited number of communication rounds, even under highly non-IID data settings. In addition, CA-FCS provides an approach that can estimate the precise carbon emissions of local training on a client device without sharing private information with the server, based on the electric charge consumed and device voltage. We conducted extensive experiments with carbon intensity data from the real world and the electricity information from three generations of Google Pixel smartphones to evaluate the effectiveness of CA-FCS under diverse datasets and data distributions. Experimental results demonstrate that CA-FCS consistently achieves substantial reductions in carbon emissions up to 46\% from client-side computation while maintaining accuracy comparable to existing state-of-the-art client selection methods. Overall, this work highlights the importance of jointly considering carbon efficiency and fairness in FL systems and provides a practical step toward more sustainable and environmentally aware FL.

\bibliographystyle{IEEEtran}
\bibliography{CarbonReferences}

\begin{thebibliography}{10}
\providecommand{\url}[1]{#1}
\csname url@samestyle\endcsname
\providecommand{\newblock}{\relax}
\providecommand{\bibinfo}[2]{#2}
\providecommand{\BIBentrySTDinterwordspacing}{\spaceskip=0pt\relax}
\providecommand{\BIBentryALTinterwordstretchfactor}{4}
\providecommand{\BIBentryALTinterwordspacing}{\spaceskip=\fontdimen2\font plus
\BIBentryALTinterwordstretchfactor\fontdimen3\font minus
  \fontdimen4\font\relax}
\providecommand{\BIBforeignlanguage}[2]{{%
\expandafter\ifx\csname l@#1\endcsname\relax
\typeout{** WARNING: IEEEtran.bst: No hyphenation pattern has been}%
\typeout{** loaded for the language `#1'. Using the pattern for}%
\typeout{** the default language instead.}%
\else
\language=\csname l@#1\endcsname
\fi
#2}}
\providecommand{\BIBdecl}{\relax}
\BIBdecl

\bibitem{wri_state_of_climate_action_2025}
\BIBentryALTinterwordspacing
C.~Schumer, S.~Boehm, J.~Jaeger, Y.~Kirana, K.~Levin, R.~Santo, K.~Lebling,
  D.~Riedl, A.~Lee, N.~Singh, M.~Sims, N.~Chin, A.~Majid, S.~Cassius, W.~Lamb,
  A.~Gangotra, N.~Grant, Y.~Zhang-Billert, and M.~Petroni, ``The state of
  climate action 2025,'' World Resources Institute, Washington, DC, Tech. Rep.,
  10 2025, report published under Systems Change Lab in collaboration with the
  Bezos Earth Fund, Climate Analytics, ClimateWorks Foundation, the Climate
  High-Level Champions, and WRI. [Online]. Available:
  \url{https://www.wri.org/research/state-climate-action-2025}
\BIBentrySTDinterwordspacing

\bibitem{towardcarbonaware}
N.~Zilberman, E.~M. Schooler, U.~Cummings, R.~Manohar, D.~Nafus, R.~Soul{\'e},
  and R.~Taylor, ``Toward carbon-aware networking,'' \emph{ACM SIGENERGY Energy
  Informatics Review}, vol.~3, no.~3, pp. 15--20, 2023.

\bibitem{carbonfootprintoffl}
X.~Qiu, T.~Parcollet, J.~Fernandez-Marques, P.~P. Gusmao, Y.~Gao, D.~J. Beutel,
  T.~Topal, A.~Mathur, and N.~D. Lane, ``A first look into the carbon footprint
  of federated learning,'' \emph{Journal of Machine Learning Research},
  vol.~24, no. 129, pp. 1--23, 2023.

\bibitem{sustainableAI1}
C.-J. Wu, R.~Raghavendra, U.~Gupta, B.~Acun, N.~Ardalani, K.~Maeng, G.~Chang,
  F.~Aga, J.~Huang, C.~Bai \emph{et~al.}, ``Sustainable ai: Environmental
  implications, challenges and opportunities,'' \emph{Proceedings of Machine
  Learning and Systems}, vol.~4, pp. 795--813, 2022.

\bibitem{sustainableAI2}
A.~Van~Wynsberghe, ``Sustainable ai: Ai for sustainability and the
  sustainability of ai,'' \emph{AI and Ethics}, vol.~1, no.~3, pp. 213--218,
  2021.

\bibitem{greenfl}
A.~Yousefpour, S.~Guo, A.~Shenoy, S.~Ghosh, P.~Stock, K.~Maeng, S.-W.
  Kr{\"u}ger, M.~Rabbat, C.-J. Wu, and I.~Mironov, ``Green federated
  learning,'' \emph{arXiv preprint arXiv:2303.14604}, 2023.

\bibitem{fedavg}
B.~McMahan, E.~Moore, D.~Ramage, S.~Hampson, and B.~A. y~Arcas,
  ``Communication-efficient learning of deep networks from decentralized
  data,'' in \emph{Artificial intelligence and statistics}.\hskip 1em plus
  0.5em minus 0.4em\relax PMLR, 2017, pp. 1273--1282.

\bibitem{poc}
Y.~J. Cho, J.~Wang, and G.~Joshi, ``Client selection in federated learning:
  Convergence analysis and power-of-choice selection strategies,'' \emph{arXiv
  preprint arXiv:2010.01243}, 2020.

\bibitem{oort}
F.~Lai, X.~Zhu, H.~V. Madhyastha, and M.~Chowdhury, ``Oort: Efficient federated
  learning via guided participant selection,'' in \emph{15th $\{$USENIX$\}$
  Symposium on Operating Systems Design and Implementation ($\{$OSDI$\}$ 21)},
  2021, pp. 19--35.

\bibitem{carbonintensity}
V.~H. Hoffmann and T.~Busch, ``Corporate carbon performance indicators: Carbon
  intensity, dependency, exposure, and risk,'' \emph{Journal of Industrial
  Ecology}, vol.~12, no.~4, pp. 505--520, 2008.

\bibitem{eafl}
A.~Arouj and A.~M. Abdelmoniem, ``Towards energy-aware federated learning on
  battery-powered clients,'' in \emph{Proceedings of the 1st ACM workshop on
  data privacy and federated learning technologies for mobile edge network},
  2022, pp. 7--12.

\bibitem{rafi2024fairness}
T.~H. Rafi, F.~A. Noor, T.~Hussain, and D.-K. Chae, ``Fairness and privacy
  preserving in federated learning: A survey,'' \emph{Information Fusion}, vol.
  105, p. 102198, 2024.

\bibitem{non-iid}
Y.~Zhao, M.~Li, L.~Lai, N.~Suda, D.~Civin, and V.~Chandra, ``Federated learning
  with non-iid data,'' \emph{arXiv preprint arXiv:1806.00582}, 2018.

\bibitem{personalizefl}
S.~Lee, O.~Tavallaie, S.~Chen, K.~Thilakarathna, S.~Seneviratne, A.~N. Toosi,
  and A.~Y. Zomaya, ``Personalizing federated learning for hierarchical edge
  networks with non-iid data,'' \emph{arXiv preprint arXiv:2504.08872}, 2025.

\bibitem{alswaitti2025training}
M.~Alswaitti, R.~Verdecchia, G.~Danoy, P.~Bouvry, and J.~Pecero, ``Training
  green ai models using elite samples,'' \emph{IEEE Transactions on Sustainable
  Computing}, 2025.

\bibitem{iftikhar2025enhancing}
S.~Iftikhar, S.~H. Alsamhi, and S.~Davy, ``Enhancing sustainability in llm
  training: Leveraging federated learning and parameter-efficient
  fine-tuning,'' \emph{IEEE Transactions on Sustainable Computing}, no.~99, pp.
  1--18, 2025.

\bibitem{sustainableAI3}
A.~G. Philipo, H.~Ning, D.~S. Sarwatt, J.~A. Mohamed, A.~S. Yusufu, F.~Shi,
  S.~Urenje, and J.~Ding, ``Sustainable ai: Emerging trends, impacts, and
  future challenges,'' \emph{IEEE Transactions on Sustainable Computing}, 2025.

\bibitem{few2024geographically}
S.~Few, P.~Djapic, G.~Strbac, J.~Nelson, and C.~Candelise, ``A geographically
  disaggregated approach to integrate low-carbon technologies across local
  electricity networks,'' \emph{Nature Energy}, vol.~9, no.~7, pp. 871--882,
  2024.

\bibitem{zivin2014spatial}
J.~S.~G. Zivin, M.~J. Kotchen, and E.~T. Mansur, ``Spatial and temporal
  heterogeneity of marginal emissions: Implications for electric cars and other
  electricity-shifting policies,'' \emph{Journal of Economic Behavior \&
  Organization}, vol. 107, pp. 248--268, 2014.

\bibitem{kairouz2021advances}
P.~Kairouz, H.~B. McMahan, B.~Avent, A.~Bellet, M.~Bennis, A.~N. Bhagoji,
  K.~Bonawitz, Z.~Charles, G.~Cormode, R.~Cummings \emph{et~al.}, ``Advances
  and open problems in federated learning,'' \emph{Foundations and
  trends{\textregistered} in machine learning}, vol.~14, no. 1--2, pp. 1--210,
  2021.

\bibitem{fedecs}
S.~Han, C.~Zhang, L.~Wang, W.~Zheng, and X.~Wen, ``Fedecs: client selection for
  optimizing computing energy in federated learning,'' in \emph{2023 IEEE 34th
  Annual International Symposium on Personal, Indoor and Mobile Radio
  Communications (PIMRC)}.\hskip 1em plus 0.5em minus 0.4em\relax IEEE, 2023,
  pp. 1--6.

\bibitem{albelaihi2022green}
R.~Albelaihi, L.~Yu, W.~D. Craft, X.~Sun, C.~Wang, and R.~Gazda, ``Green
  federated learning via energy-aware client selection,'' in \emph{GLOBECOM
  2022-2022 IEEE Global Communications Conference}.\hskip 1em plus 0.5em minus
  0.4em\relax IEEE, 2022, pp. 13--18.

\bibitem{cefl}
T.~Mehboob, N.~Bashir, J.~O. Iglesias, M.~Zink, and D.~Irwin, ``Cefl:
  Carbon-efficient federated learning,'' \emph{arXiv preprint
  arXiv:2310.17972}, 2023.

\bibitem{cafe}
J.~Bian, L.~Wang, S.~Ren, and J.~Xu, ``Cafe: Carbon-aware federated learning in
  geographically distributed data centers,'' in \emph{Proceedings of the 15th
  ACM International Conference on Future and Sustainable Energy Systems}, 2024,
  pp. 347--360.

\bibitem{khan2018analysis}
I.~Khan, M.~W. Jack, and J.~Stephenson, ``Analysis of greenhouse gas emissions
  in electricity systems using time-varying carbon intensity,'' \emph{Journal
  of Cleaner Production}, vol. 184, pp. 1091--1101, 2018.

\bibitem{gao2024federated}
W.~Gao, O.~Tavallaie, S.~Chen, and A.~Zomaya, ``Federated learning as a service
  for hierarchical edge networks with heterogeneous models,'' in
  \emph{International Conference on Service-Oriented Computing}.\hskip 1em plus
  0.5em minus 0.4em\relax Springer, 2024, pp. 85--99.

\bibitem{esquivel2024performance}
J.~A. Esquivel, B.~Aldous, A.~M. Abdelmoniem, A.~Alhilal, and L.~You,
  ``Performance profiling of federated learning across heterogeneous mobile
  devices,'' in \emph{2024 IEEE 24th International Conference on Software
  Quality, Reliability, and Security Companion (QRS-C)}.\hskip 1em plus 0.5em
  minus 0.4em\relax IEEE, 2024, pp. 363--372.

\bibitem{fairness}
Y.~Shi, Z.~Liu, Z.~Shi, and H.~Yu, ``Fairness-aware client selection for
  federated learning,'' in \emph{2023 IEEE international conference on
  multimedia and expo (ICME)}.\hskip 1em plus 0.5em minus 0.4em\relax IEEE,
  2023, pp. 324--329.

\bibitem{aryai2024real}
V.~Aryai and M.~Goldsworthy, ``Real-time high-resolution modelling of grid
  carbon emissions intensity,'' \emph{Sustainable Cities and Society}, vol.
  104, p. 105316, 2024.

\bibitem{arputharaj2025green}
D.~R. Arputharaj, C.~Rodriguez, A.~Rodio, and G.~Neglia, ``Green federated
  learning via carbon-aware client and time slot scheduling,'' in \emph{2025
  33rd International Symposium on Modeling, Analysis and Simulation of Computer
  and Telecommunication Systems (MASCOTS)}.\hskip 1em plus 0.5em minus
  0.4em\relax IEEE, 2025, pp. 1--8.

\bibitem{yang2025survey}
J.~Yang, Z.~Saad, J.~Wu, X.~Niu, H.~Leung, and S.~Drew, ``A survey on task
  scheduling in carbon-aware container orchestration,'' \emph{arXiv preprint
  arXiv:2508.05949}, 2025.

\bibitem{electricitymaps}
\BIBentryALTinterwordspacing
{Electricity Maps}. (2025) Electricity maps -- south australia zone data (72h,
  hourly). Accessed: Sep. 2, 2025. [Online]. Available:
  \url{https://app.electricitymaps.com/zone/AU-SA/72h/hourly}
\BIBentrySTDinterwordspacing

\bibitem{hasan2025carbon}
S.~M. Hasan, T.~Islam, M.~Saifuzzaman, K.~R. Ahmed, C.-H. Huang, and A.~R.
  Shahid, ``Carbon emission quantification of machine learning: A review,''
  \emph{IEEE Transactions on Sustainable Computing}, 2025.

\bibitem{mnist}
Y.~LeCun, C.~Cortes, and C.~J.~C. Burges, ``The mnist database of handwritten
  digits,'' 1998, available: \url{http://yann.lecun.com/exdb/mnist/}.

\bibitem{fmnist}
H.~Xiao, K.~Rasul, and R.~Vollgraf, ``Fashion-{MNIST}: a novel image dataset
  for benchmarking machine learning algorithms,'' 2017.

\bibitem{cifar}
A.~Krizhevsky, V.~Nair, and G.~Hinton, ``Learning multiple layers of features
  from tiny images,'' University of Toronto, Tech. Rep., 2009, technical
  Report.

\bibitem{cinic10}
L.~N. Darlow, E.~J. Crowley, A.~Antoniou, and A.~J. Storkey, ``{CINIC-10} is
  not {ImageNet} or {CIFAR-10},'' 2018.

\bibitem{imagenet}
J.~Deng, W.~Dong, R.~Socher, L.-J. Li, K.~Li, and L.~Fei-Fei, ``Imagenet: A
  large-scale hierarchical image database,'' in \emph{Proceedings of the IEEE
  Conference on Computer Vision and Pattern Recognition (CVPR)}.\hskip 1em plus
  0.5em minus 0.4em\relax IEEE, 2009, pp. 248--255.

\bibitem{fedtctf}
C.~Liang, W.~Liang, Y.~Chen, K.~Li, and A.~Y. Zomaya, ``Fedtctf: Tensor
  completion-based federated learning for device heterogeneity,'' \emph{IEEE
  Transactions on Sustainable Computing}, 2025.

\bibitem{pixel6a_recall}
\BIBentryALTinterwordspacing
{Australian Competition and Consumer Commission}. (2025) Product recall notice:
  Google pixel 6a smartphone. Accessed: Sep. 5, 2025. [Online]. Available:
  \url{https://www.productsafety.gov.au/search-consumer-product-recalls/google-pixel-6a-smartphone}
\BIBentrySTDinterwordspacing

\bibitem{adb}
\BIBentryALTinterwordspacing
{Android Developers}. (2025) Android debug bridge (adb). Accessed: Sep. 5,
  2025. [Online]. Available: \url{https://developer.android.com/tools/adb}
\BIBentrySTDinterwordspacing

\bibitem{mammen2021wisleep}
P.~M. Mammen, C.~Zakaria, T.~Molom-Ochir, A.~Trivedi, P.~Shenoy, and R.~Balan,
  ``Wisleep: Inferring sleep duration at scale using passive wifi sensing,''
  \emph{arXiv preprint arXiv:2102.03690}, 2021.

\end{thebibliography}

\begin{IEEEbiography}
[{\includegraphics[width=1in,height=1.25in,clip,keepaspectratio]{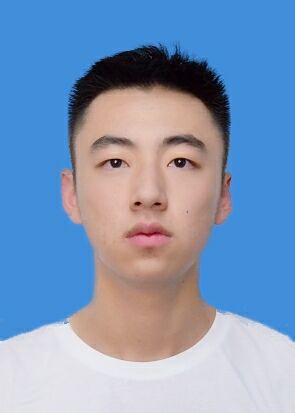}}]
{Wentao Gao} received the  Bachelor of Advanced
Computing degree (Hons.) in computational data science from the University of Sydney, Camperdown, NSW, Australia, in 2023. He is currently pursuing the Ph.D. degree in computer science at the University of Sydney, Sydney, Australia. His research interests include sustainable computing, federated learning, and distributed systems.
\end{IEEEbiography}

\begin{IEEEbiography}
[{\includegraphics[width=1in,height=1.25in,clip,keepaspectratio]{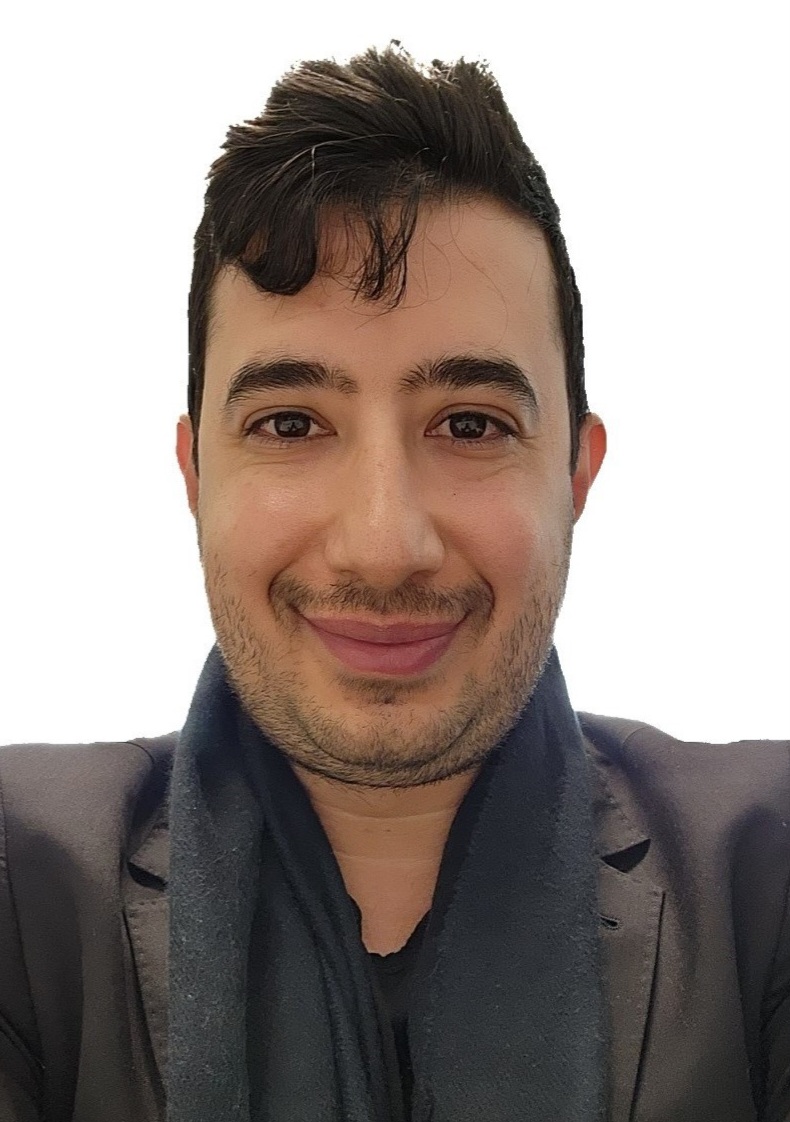}}]
{Omid Tavallaie} is a Lecturer in the Department of Computer Science at the University of Western Australia and a Visiting Academic in the Department of Engineering Science at the University of Oxford. Before joining UWA, he held postdoctoral research positions at both the University of Oxford and the University of Sydney. He earned his PhD in Computer Science from the University of Sydney in 2021. Between 2021 and 2024, he also lectured postgraduate courses at the University of Sydney. Dr Tavallaie has over seven years of teaching experience as a lecturer. His research focuses on sustainable computer networking, distributed systems, the Internet of Things (IoT), and federated learning.
\end{IEEEbiography}

\begin{IEEEbiography}
[{\includegraphics[width=1in,height=1.25in,clip,keepaspectratio]{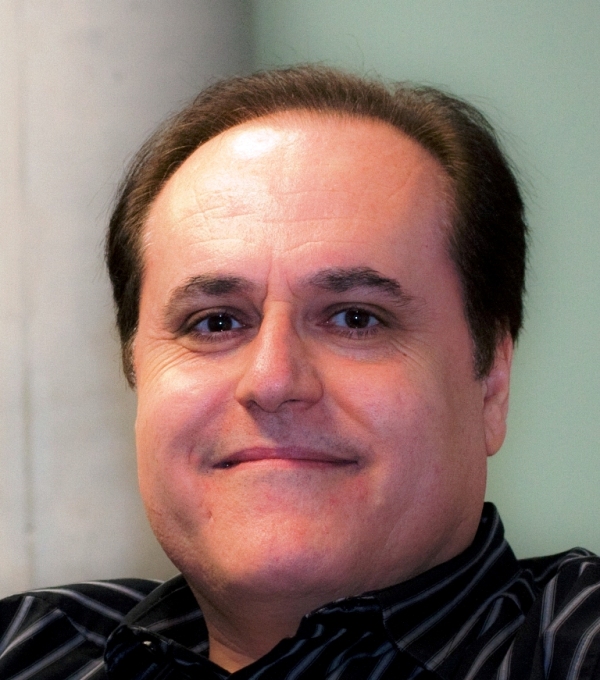}}]
{Albert Y. Zomaya} is the Peter Nicol Russell Chair Professor of Computer Science at the University of Sydney. A global leader in parallel and distributed systems, he has authored more than 800 publications and 30 books, shaping the field’s research agenda for over three decades. He is a Fellow of the IEEE, the Australian Academy of Science, and the Royal Society of New South Wales, and an elected member of Academia Europaea and the European Academy of Sciences and Arts. Professor Zomaya previously served as Editor-in-Chief of IEEE Transactions on Computers, IEEE Transactions on Sustainable Computing, and ACM Computing Surveys. His work spans parallel and distributed computing, networking, and complex systems, with a lasting influence on both theory and practice.
\end{IEEEbiography}

\vfill

\end{document}